\documentclass[aps,amsmath,amssymb,superscriptaddress,reprint]{revtex4-2}
\usepackage[utf8]{inputenc}
\usepackage[T1]{fontenc}
\usepackage{etoolbox}
\usepackage{graphicx}
\usepackage{dcolumn}
\usepackage{bm}
\usepackage{blindtext}
\usepackage{physics}
\usepackage{siunitx}
\usepackage{float}
\usepackage{xcolor}
\usepackage[acronym]{glossaries}
\usepackage{soul}

\newcommand{\liouv}{i\mathcal{L}}

\newcommand{\QQ}{\mathcal{Q}}
\newcommand{\MM}{\nu}
\newcommand{\PP}{\mathcal{P}}

\newcommand{\ttiny}[1]{\text{\tiny{#1}}}
\newcommand{\w}[0]{\omega}

\DeclareMathOperator*{\argmin}{arg\,min}

\definecolor{UniBlue}{HTML}{5F6DA3}
\definecolor{TitleBlue}{HTML}{EBEDF5}
\definecolor{TxtBlue}{HTML}{182D80}
\definecolor{TxtRed}{HTML}{CC3300}
\definecolor{TxtGreen}{HTML}{339900}
\definecolor{SecBlue}{HTML}{434E76}
\definecolor{TxtBlack}{HTML}{212121}
\definecolor{TxtGrey}{HTML}{828282}

\begin{document}

\title[]{Routine Molecular Dynamics Simulations Including Nuclear Quantum Effects: from Force Fields to Machine Learning Potentials}
\author{Thomas Plé*}
\affiliation{Sorbonne Université, LCT, UMR 7616 CNRS, F-75005, Paris, France}
\author{Nastasia Mauger}
\affiliation{Sorbonne Université, LCT, UMR 7616 CNRS, F-75005, Paris, France}
\author{Olivier Adjoua}
\affiliation{Sorbonne Université, LCT, UMR 7616 CNRS, F-75005, Paris, France}
\author{Théo Jaffrelot-Inizan}
\affiliation{Sorbonne Université, LCT, UMR 7616 CNRS, F-75005, Paris, France}
\author{Louis Lagardère*}
\affiliation{Sorbonne Université, LCT, UMR 7616 CNRS, F-75005, Paris, France}
\author{Simon Huppert}
\affiliation{Institut des Nanosciences de Paris (INSP), CNRS UMR 7588 and Sorbonne Université, F-75005, Paris, France}
\author{Jean-Philip Piquemal*}
\affiliation{Sorbonne Université, LCT, UMR 7616 CNRS, F-75005, Paris, France}
\email{jean-philip.piquemal@sorbonne-université.fr}


\begin{abstract}
We report the implementation of a multi-CPU and multi-GPU massively parallel platform dedicated to the explicit inclusion of nuclear quantum effects (NQEs) in the Tinker-HP molecular dynamics (MD) package. The platform, denoted Quantum-HP, exploits two simulation strategies: the Ring-Polymer Molecular Dynamics (RPMD) that provides exact structural properties at the cost of a MD simulation in an extended space of multiple replicas, and the adaptive Quantum Thermal Bath (adQTB) that imposes the quantum distribution of energy on a classical system via a generalized Langevin thermostat and provides computationally affordable and accurate (though approximate) NQEs.
We discuss some implementation details, efficient numerical schemes, parallelization strategies and quickly review the GPU acceleration of our code. Our implementation allows an efficient inclusion of NQEs in MD simulations for very large systems, as demonstrated by scaling tests on water boxes with more than 200,000 atoms (simulated using the AMOEBA polarizable force field). We test the compatibility of the approach with Tinker-HP's recently introduced Deep-HP machine learning potentials module by computing water properties using the DeePMD potential with adQTB thermostating. Finally, we show that the platform is also compatible with the alchemical free energy estimation capabilities of Tinker-HP and fast enough to perform simulations. Therefore, we study how the NQEs affect the hydration free energy of small molecules solvated with the recently developed Q-AMOEBA water force field. Overall, the Quantum-HP platform allows users to perform routine quantum MD simulations of large condensed-phase systems and will participate to shed a new light on the quantum nature of important interactions in biological matter. 
\end{abstract}

\maketitle


\section{Introduction}\label{sec:introduction}

Molecular dynamics (MD) is a powerful simulation tool that allows to compute properties of atomistic systems in a wide range of conditions, with the aim of explaining experimental results or even be predictive. Over the last decades, it has been a very active field of research. 
Long and accurate simulations of large condensed-phase systems are now reachable with recent advances in High Performance Computing (HPC) and GPU acceleration. We can distinguish efforts made in this field in two categories: a) improvements of the models for interatomic interactions, b) more efficient and accurate simulation of the nuclear motion in the desired statistical ensemble. Regarding the first category, considerable improvements have been made in two directions: efficiency of first principle descriptions (for example using Born-Oppenheimer Density Functional Theory) on the one hand, and accuracy of effective models (classical force fields\cite{case2005amber,brooks2009charmm,van2005gromacs}, polarizable force fields\cite{chapterpol2015,reddy2016accuracy,Meclr19,annurev-biophys}, "machine-learning" (ML) force fields\cite{behler2007generalized,zhang2018deep,smith2017ani}) on the other hand.

Regarding the second category, lots of attention has been given to the development of efficient integration schemes (multi-timestepping~\cite{tuckerman1992reversible,lagardere2019pushing}, hybrid Monte Carlo algorithms~\cite{duane1987hybrid,girolami2011riemann},...) or improved sampling methods (parallel tempering~\cite{sugita1999replica}, metadynamics~\cite{laio2002escaping},...) in order to tackle the need for long simulations in complex energy landscapes. Most implementations however, assume that the nuclei are classical particles, thus completely neglecting nuclear quantum effects (NQEs) or implicitly including them in an uncontrolled manner -- for example by fitting force fields on experimental data simulated using classical MD -- which limits transferability~\cite{paesani2007quantum,fanourgakis2006quantitative,li2022static}.

As MD simulations grow in accuracy and efficiency, the need for the explicit inclusion of NQEs becomes more and more apparent, be it in simulations of systems in extreme conditions (low temperatures, high pressures) where they can be massive~\cite{benoit1998tunnelling,schaack2020quantum,brieuc2020converged}, or even in more standard conditions where it has already been shown that more subtle NQEs are at play\cite{paesani2009properties,schwartz2009importance,ceriotti2013nuclear,engel2021importance}. 
NQEs can be explicitly included in MD simulations in the framework of path integrals (PIMD) which provides an exact description of structural NQEs~\cite{feynman2010quantum,berne1986simulation}. 
Even though they are considered as the gold standard, PIMD calculations are usually expensive as they require to simulate the system in an extended phase space which size grows when NQEs are more pronounced.
Cheaper approximate methods have been recently developed~\cite{Ceriotti_PRL2009,dammak2009quantum,ceriotti2011accelerating,brieuc2016quantum}, among which the adaptive quantum thermal bath (adQTB)~\cite{mangaud2019fluctuation,huppert2022simulation} that proved to be an accurate alternative to PIMD at the cost of a classical MD simulation~\cite{mauger2021nuclear}.
As NQEs are suspected to play a role in some biological processes~\cite{fang2016inverse,law2018importance}, an efficient and parallel implementation of these methods is required to simulate the large systems and long timescales involved in such processes.
As highlighted in several previous papers~\cite{paesani2007quantum,pereyaslavets2018importance,mauger2022improving}, it is also desirable to design advanced force fields (FFs) with explicit NQEs from scratch (\textit{i.e.} that do not implicitly incorporate them through parametrization).  This endeavour, which is already challenging in a classical framework, was up to now nearly unachievable as it requires numerous quantum simulations to adapt the parameters of the model, and because highly efficient implementations of PIMD or adQTB in standard MD codes are scarce.

In this work, we report the implementation of Quantum-HP, a highly-parallel platform for the explicit inclusion of NQEs, compatible with multi-GPU acceleration, inside the Tinker-HP molecular dynamics package~\cite{lagardere2018tinker,adjoua2021tinker}. The platform is fully compatible with all the force fields present in Tinker-HP, including classical ones (CHARMM, AMBER) and the AMOEBA~\cite{ren2003polarizable,ren2011polarizable,shi2013polarizable}, AMOEBA+~\cite{liu2019amoebaplus,liu2019implementation} and SIBFA~\cite{Gresh2007, naseem2022development} polarizable FFs, and allows the simulation of million-atoms systems in a distributed architecture. The paper is organized as follows: section~\ref{sec:methods} briefly describes the theory for the two methods that we implemented, namely Ring-Polymer MD and the adQTB. Section~\ref{sec:implementation} provides some important implementation details for both methods, including time integrators and parallelization strategies. Scaling and efficiency tests are also provided, as well as a brief description of the GPU acceleration. We briefly show in section~\ref{sec:deephp} that Quantum-HP is compatible with the new Deep-HP~\cite{inizan2022scalable} platform that allows to perform molecular dynamics using machine-learning (ML) potentials or hybrid ML/MM force fields. Finally, in section~\ref{sec:applications} we demonstrate the capabilities of the platform and the accuracy of the recently developed Q-AMOEBA force field~\cite{mauger2022improving} by computing the hydration free energies of a benchmark dataset of small organic molecules, for which we obtain state-of-the-art accuracy when including NQEs using the adQTB. Section \ref{sec:conclusion} provides some concluding remarks and outlooks for future developments and applications. 

\section{Methods}\label{sec:methods}
In this section, we will briefly describe the theoretical framework of the two methods for the inclusion of NQEs, namely Ring-Polymer Molecular Dynamics (RPMD) and the adaptive Quantum Thermal Bath (adQTB), that we implemented in Tinker-HP.

\subsection{Ring-Polymer Molecular Dynamics}
\begin{figure}
    \centering
    \includegraphics{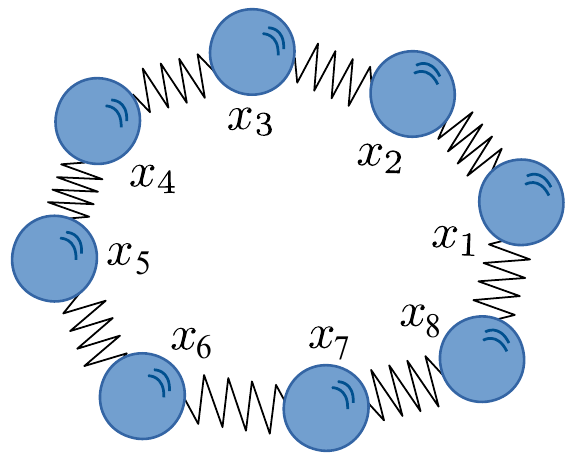}
    \caption{Schematic representation of the ring-polymer path-integral for $\MM=8$. Each bead  $x_1,\hdots,x_\MM$ (represented by a blue circle) is subject to the physical potential and connected to its nearest neighbours via a harmonic potential (represented as springs).}
    \label{fig:closed_chain_schema}
\end{figure}

Ring-Polymer Molecular Dynamics~\cite{Craig_manolopoulosJCP2004,habershon2013ring} is based on the imaginary-time path integral formulation of quantum statistical mechanics. This formalism allows to express the canonical partition function $Z=Tr\qty[e^{-\beta\hat{H}}]$ of a quantum system (made of distinguishable particles at thermal equilibrium)
as the one of an effective classical system. This system takes the form of a so-called "ring polymer" (as schematically depicted in Figure~\ref{fig:closed_chain_schema}) where "beads" along the polymer are replicas of the whole original system (independently subject to the interatomic potential $V$) that interact through a harmonic potential. In particular, in our implementation, we employ the scaled normal modes representation of the ring-polymer that describes it in terms of a center of mass (called the centroid) and fluctuations around it. In this framework, 
the quantum partition function is written as:
\begin{equation}\label{eq:isomorphism_RPMD}
     Z=\lim_{\MM\to\infty}\int \dd\QQ~e^{-\beta\qty(U_\MM(\QQ) ~+~ \sum_{n>0} \frac{1}{2}\omega_n^2Q_n^T M Q_n)}
\end{equation}
where $\QQ=(Q_0,\hdots,Q_{\MM-1})$ are the amplitudes of the $\MM$ modes describing the ring-polymer (each being a vector of size $3N_{atoms}$), $M$ is the diagonal mass matrix of the physical system and $\omega_n$ are the characteristic frequencies of the normal modes which are defined as the square roots of the eigenvalues (ordered by increasing amplitude) of the $\MM\times\MM$ matrix:
\begin{equation}\label{eq:matrix_free_RP}
    \Omega_\MM^2=\frac{\MM^2}{\hbar^2\beta^2}
    \begin{pmatrix}
        2  & -1 &  &  & -1 \\
        -1 & \ddots  & \ddots &   &\\
           & \ddots & \ddots& \ddots& & \\
         &  & \ddots & \ddots& -1 &   \\
        -1 &  &  & -1 & 2 
    \end{pmatrix}
\end{equation}
where all the undefined terms in the matrix are zeros. The normal modes are subject to the potential $U_\MM(\QQ)$ which is defined as:
\begin{equation}\label{eq:PI_potential}
    U_\MM(\QQ)=\frac{1}{\MM}\sum_{i=0}^{\MM-1} V\qty(x_i^{(\MM)}(\QQ))
\end{equation}
where $V$ is the physical interatomic potential and $x_i^{(\MM)}$ is the position of the $i$th bead of the ring-polymer, that is constructed from the $\MM$ normal mode amplitudes as:
\begin{equation}\label{eq:beads_positions}
    x_i^{(\MM)}(\QQ) = Q_0 + \sqrt{\MM}\sum_{n=1}^{\MM-1} T^{(\MM)}_{in} Q_n
\end{equation}
with $T^{(\MM)}$ the unitary transfer matrix which columns are the eigenvectors of the matrix \eqref{eq:matrix_free_RP}. We note that $Q_0$ represents the position of the centroid of the ring polymer (with associated frequency $\w_0=0$) and that $Q_{n>0}$ are called fluctuation modes. From eq.~\eqref{eq:isomorphism_RPMD}, we write the probability distribution of the ring-polymer $\rho_\MM(\QQ)$ as:
\begin{equation}
    \rho_\MM(\QQ)=\frac{1}{Z_\MM}e^{-\beta\qty(U_\MM(\QQ) ~+~ \sum_{n>0} \frac{1}{2}\omega_n^2Q_n^T M Q_n)}
\end{equation}
with $Z_\MM$ a normalization constant such that $Z~=~\lim_{\MM\to\infty}~Z_\MM$.
In this framework, the thermal equilibrium average of any position-dependent observable $A(\hat x)$ is obtained as an average over the distribution $\rho_\MM$ (in the limit $\MM\to\infty$):
\begin{equation}\label{eq:expval_RPMD}
    \expval{A(\hat{x})}_\beta =\Tr\qty[A(\hat x)\frac{e^{-\beta\hat{H}}}{Z}]= \lim_{\MM\to\infty}\int \dd \QQ~A_\MM(\QQ)~\rho_\MM(\QQ)
\end{equation}
with $A_\MM(\QQ)=\sum_{i=0}^{\MM-1} A\left(x_i^{(\MM)}(\QQ)\right)/\MM$ defined similarly as in eq.~\eqref{eq:PI_potential} for the potential energy.

In order to perform molecular dynamics simulations, a set of momenta $\PP=(P_0,\hdots,P_{\MM-1})$ are associated to the normal modes so that the joint probability density becomes:
\begin{equation}\label{eq:distrib_RPMD}
    \rho_\MM(\QQ,\PP)\propto\rho_\MM(\QQ)~e^{-\beta\sum_{n}\frac{1}{2}P_n^T M^{-1} P_n}
\end{equation}
This formalism also allows to compute approximate (Kubo-transformed) time correlation functions of position-dependent observables as~\cite{hele2017thermal,althorpe2021path}:
\begin{equation}\label{eq:Kab_RPMD}
    K_{AB}^{(\MM)}(t)=\int \dd \QQ \dd\PP~\rho_\MM(\QQ,\PP)~A_\MM(\QQ) B_\MM(\QQ(t))
\end{equation}
where $\QQ(t)$ is obtained by propagating for a duration $t$ the ring-polymer equations of motion:
\begin{equation}\label{eq:eom_rpmd}
    \left\{~
    \begin{aligned}
    \dot{Q_n}&= M^{-1}P_n\\
    \dot{P_n}&=f_n(\QQ)- \omega_n^2 M Q_n\\
    \end{aligned}
    \right. \quad (n=0,\hdots,\MM-1)
\end{equation}
with $f_n(\QQ)$ the interatomic force projected on the $n$th normal mode which is obtained from the chain rule as:
\begin{align}
    f_0(\QQ)=&-\frac{1}{\MM}\sum_{i=0}^{\MM-1}\nabla V(x_i^{(\MM)}(\QQ))\\
    f_{n>0}(\QQ)=&-\frac{1}{\sqrt{\MM}}\sum_{i=0}^{\MM-1}T_{in}^{(\MM)}\nabla V(x_i^{(\MM)}(\QQ)) \label{eq:force_normal_modes}
\end{align}
Importantly, we note that while equation \eqref{eq:expval_RPMD} is exact in the $\MM\to\infty$ limit independently of the form of $V$, this is not the case for eq.~\eqref{eq:Kab_RPMD} as the dynamics of the ring-polymer generated by the equations of motion \eqref{eq:eom_rpmd} does not generally reproduce the exact quantum dynamics~\cite{braams2006short,hele2015communication}. This approximation was however shown to be quite robust and to provide relevant results in many applications~\cite{witt2009applicability,habershon2013ring,benson2020quantum}.

\subsection{Adaptive Quantum Thermal Bath}
 The adaptive Quantum Thermal Bath (adQTB) is an hybrid quantum-classical method that relies on a generalized Langevin thermostat in order to impose the quantum distribution of energy on a classical system~\cite{dammak2009quantum,mangaud2019fluctuation}:
\begin{equation}\label{eq_langevin}
    \left\{~
    \begin{aligned}
    \dot{x}&= M^{-1}p\\
    \dot{p}&=-\nabla V(x)-\gamma_0 p+F(t)\\
    \end{aligned}
    \right.
\end{equation}
where $\gamma_0$ is a friction coefficient and $F(t)$ is a colored random force with the following power spectrum:
\begin{equation}\label{eq:Cff_qtb}
    C_{F_i F_j}(\omega)=2m_i\gamma_i(\omega)\Theta(\omega,\beta)\delta_i^j \qquad (i,j = 1,\hdots,3N_{atoms})
\end{equation}
with $m_i$ the $i$th diagonal element of the mass matrix $M$, $\delta_i^j$ the Kroneker delta symbol and
\begin{equation}
    \Theta(\omega,\beta) = \frac{\hbar\omega/2}{\tanh(\beta\hbar\omega/2)}
\end{equation}
the average thermal energy of a quantum harmonic oscillator of frequency $\omega$ at inverse temperature $\beta$. The parameters $\gamma_i(\omega)$ in the random force amplitude are adjusted in order to minimize the average deviation from the quantum Fluctuation-Dissipation Theorem~\cite{kubo1966fluctuation,mangaud2019fluctuation,huppert2022simulation} (FDT):
\begin{align}
    \gamma_i^\star(\w) &=\argmin_{\gamma_i(\w)}\abs{\Delta_\ttiny{FDT,i}(\omega)} \nonumber\\ &=\argmin_{\gamma_i(\w)}\abs{m_i\gamma_i(\w)C_{v_iv_i}(\w)-\Re[C_{v_iF_i}(\w)]}
\end{align}
where $C_{v_iv_i}(\w)$ (resp. $C_{v_iF_i}(\w)$) is the velocity-velocity (resp. velocity-random force) correlation spectrum estimated in a QTB simulation using the trial parameter $\gamma_i$ in the random force power spectrum. The optimum $\Delta_\ttiny{FDT,i}(\omega)=0$ indicates that the thermal energy (including Zero-Point Energy) is correctly distributed in the system, according to the quantum FDT. 

For a purely harmonic system, $\gamma_i^\star(\w)$ is known analytically and one can show that $\Delta_\ttiny{FDT,i}(\w)=0$ for constant $\gamma_i^\star(\w)=\gamma_0, \forall \omega$. Additionally, in this particular case, the QTB dynamics produces the exact quantum phase space distribution for sufficiently small values of $\gamma_0$~\cite{Basire2013}. The original QTB, as devised in ref.~\cite{dammak2009quantum}, is obtained by using the harmonic solution $\gamma_i(\w)=\gamma_0$, even for anharmonic systems. Deviations from the quantum FDT might therefore be present, that manifest through the well-documented ZPE leakage~\cite{brieuc2016_ZPEL,habershon2009zero}.

The fluctuation-dissipation theorem 
provides a generic criterion to optimize the parameters for any anharmonic system and no \textit{a priori} information on the system is required. Deviations $\Delta_\ttiny{FDT,i}(\w)$ from the quantum FDT are estimated along the dynamics 
and used to adapt the adQTB parameters $\gamma_i(\w)$ with a procedure detailed in section~\ref{sec:qtb_spectra_adapt}. In practice, the estimator of $\Delta_\ttiny{FDT,i}(\w)$ is subject to statistical noise so that we do not strictly optimize the parameters but rather let them fluctuate around their optimal value such that $\Delta_\ttiny{FDT,i}(\w)$ should fluctuate around zero.
The adaptation procedure is performed in an equilibration phase which duration typically ranges from a few picoseconds to a few hundred picoseconds depending on the system.

While the adQTB cannot be formally derived from first principles, it was recently shown to provide accurate results even in very anharmonic systems~\cite{mauger2021nuclear}. As its computational cost is essentially the same as that of a classical MD simulation, it is a promising approach for the quantum simulation of large biological systems. It was the method of choice for the recent development of the Q-AMOEBA force field~\cite{mauger2022improving} and we will show in section \ref{sec:applications} that the combination Q-AMOEBA/adQTB can be used to accurately compute hydration free energies of small organic molecules.


\section{Implementation}\label{sec:implementation}

This section provides implementation details for both methods, starting with RPMD and following with adQTB. Integration schemes and parallelization strategies are discussed, as well as some technical points specific to each method. To conclude this section, some scaling tests (using the AMOEBA polarizable FF) are presented in order to compare the efficiency of the quantum methods compared to reference classical MD calculations.

\subsection{RPMD}\label{subsec:implementation_rpmd}
\subsubsection{Integration Scheme}~\label{sec:RPMD_integration}
In order to sample the canonical distribution of the ring-polymer \eqref{eq:distrib_RPMD}, we attach a Langevin thermostat to each normal mode. The equations of motion are then integrated using the BAOAB scheme, originally introduced by Leimkhuller \textit{et al.}~\cite{Leimkuhler2012} and adapted for path-integrals simulations (following for example ref.~\cite{LiuLiLiu}). The choice of the normal mode representation allows to efficiently integrate the rapidly oscillating motion due to the path-integral harmonic chain and to use a simulation timestep that is essentially dictated by the characteristic timescales of the interatomic potential $V$ (and is thus similar to classical simulations). Our implementation also utilizes the TRPMD scheme of ref.~\cite{Rossi2014} in which we apply a strong (critically damped) Langevin thermostat to the fluctuation modes. In order to ensure ergodicity while minimizing the disruption to the dynamics, we apply an underdamped Langevin thermostat to the centroid (the original TRPMD is thus recovered in the limit of zero damping on the centroid). The TRPMD equations of motion read as follows:
\begin{multline}\label{eq:RPMD_EOM}
    \begin{pmatrix}
        \dot{Q_n}\\
        \dot{P_n}
    \end{pmatrix}
    = \overbrace{\begin{pmatrix}
        M^{-1}P_n\\
        -\omega_n^2 M Q_n
    \end{pmatrix}}^{e^{\liouv_A t}}
    +\overbrace{\begin{pmatrix}
        0\\
        f_n(\QQ)
    \end{pmatrix}}^{e^{\liouv_B t}}\\
    +\underbrace{\begin{pmatrix}
        0\\
        -\gamma_n P_n + (2\gamma_n\beta^{-1}M)^\frac{1}{2}R_n(t)
    \end{pmatrix}}_{e^{\liouv_O t}}
\end{multline}
with $R_n(t)$ a $3N_{atoms}$ vector of uncorrelated standard Gaussian white noise and $\gamma_n=\max(\gamma_0,\omega_n)$ the (critical) friction coefficient for each normal mode.
We decompose the equations of motion into three analytically solvable blocks which formal solutions are denoted by the corresponding Liouville propagators $e^{\liouv_A t}$,$e^{\liouv_B t}$ and $e^{\liouv_O t}$. The propagator $e^{\liouv_A t}$ corresponds to a harmonic evolution of the fluctuation modes and a simple translation of the centroid position (since $\omega_0=0$). The propagator $e^{\liouv_B t}$ is a translation of the momenta according to the interatomic forces (projected on the normal modes) and the propagator $e^{\liouv_O t}$ is a standard Ornstein-Uhlenbeck process~\cite{gillespie1996exact} for each normal mode.
The full time propagator over a duration $t=n_{step}\Delta t$ is then symmetrically broken up as:
\begin{equation}
    e^{\liouv_\ttiny{RPMD}t}\approx\qty(
    e^{\liouv_{B}\frac{\Delta t}{2}}e^{\liouv_A\frac{\Delta t}{2}}e^{\liouv_O\Delta t}e^{\liouv_A\frac{\Delta t}{2}}e^{\liouv_{B}\frac{\Delta t}{2}}
    )^{n_{step}}
\end{equation}
where $\Delta t$ is the simulation timestep. 
The implementation also optionally allows the use of the BCOCB variant recently introduced in ref.~\cite{korol2020dimension} where the exact integration of $e^{\liouv_A\frac{\Delta t}{2}}$ is replaced by a numerical scheme that allows for a better stability of the dynamics and larger timesteps (a three-fold increase in some cases) when a large number of beads is required.

\subsubsection{Multi-timestep Methods}
\label{subsec:multistep}
For interatomic potentials of the form:
\begin{equation}
\label{eq:splitting_sf}
    V={V_s}+{V_f}
\end{equation}
with $V_s$ a slowly varying and expensive component of the total potential (non-bonded interactions in the case of AMOEBA) and $V_f$ a quickly varying and inexpensive component (bonded interactions in the case of AMOEBA), simulations can be made more efficient using multiple timestepping 
to compute the expensive $V_s$ less frequently. In the case of RPMD, the splitting \eqref{eq:splitting_sf} can be used to our advantage both in the time integration scheme (RESPA algorithm~\cite{tuckerman1992reversible}) and in the computation of the interatomic forces on the ring-polymer beads (ring-polymer contraction).\\

\paragraph{Path-integral RESPA integrator}
~\\

To improve the efficiency of the integration scheme, we use the BAOAB-RESPA multi-timestep algorithm, which was initially designed for classical Langevin MD~\cite{tuckerman1992reversible,lagardere2019pushing} and adapted to path-integrals simulations~\cite{tuckerman1993efficient,martyna1996explicit,martyna1999molecular}. 
For this method, the $B$ block (update of the velocities according to the interatomic forces) is split into $Bf$ (associated to $\nabla V_f$) and $Bs$ block (associated to  $\nabla V_s$) and the dynamics is propagated using a two-stage symmetric Trotter break-up of the full time-propagator:
\begin{multline}
    e^{\liouv_\ttiny{RPMD} t}\approx\Bigg(
    {e^{\liouv_{Bs}\frac{n_{alt}\Delta t}{2}}}\\
    \qty(
    {e^{\liouv_{Bf}\frac{\Delta t}{2}}}e^{\liouv_A\frac{\Delta t}{2}}e^{\liouv_O\Delta t}e^{\liouv_A\frac{\Delta t}{2}}{e^{\liouv_{Bf}\frac{\Delta t}{2}}}
    )^{n_{alt}}\\
    {e^{\liouv_{Bs}\frac{n_{alt}\Delta t}{2}}}
    \Bigg)^{n_{step}}
\end{multline}
This expression shows that, for a full time step of integration, an inner loop of $n_{alt}$ BfAOABf timesteps is performed with a short timestep $\Delta t$. Once every $n_{alt}$ timestep, a propagation of the block $Bs$ is performed with the larger timestep $n_{alt}\Delta t$. When the computation of $\nabla V_s$ dominates the total calculation time, this scheme allows performance gains of up to a factor $n_{alt}$. Typically, the smaller timestep $\Delta t$ ranges between 0.2fs to 1fs while the larger timestep $n_{alt}\Delta t$ is of the order of 2fs.\\

\paragraph{Ring-polymer Contractions}
~\\

Taking further advantage of the separation of the interatomic potential $V={V_s}+{V_f}$ used in the RESPA integrator, we implemented the Ring-polymer contraction (RPC) scheme introduced in refs.~\cite{markland2008efficient,markland2008refined}. This scheme is based on the assumption that the motion of high-frequency normal modes of the ring-polymer is only weakly affected by the slowly varying inter-atomic forces, so that one can neglect these modes when evaluating the slow forces. This allows to evaluate the slowly-varying potential on a "contracted" set of beads instead of the full ring-polymer:
\begin{equation}
    U_\MM(\QQ)\approx \frac{1}{\MM}\sum_{i=0}^{\MM-1}V_f(x_i^{(\MM)}(\QQ))
        + \frac{1}{\tilde{\MM}}\sum_{i=0}^{\tilde{\MM}-1}V_s(x_i^{(\tilde{\MM})}(\QQ))
\end{equation}
with $\tilde{\MM}\leq\MM$ and where the coordinates $x_i^{(\tilde{\MM})}$ are computed similarly as in eq.~\eqref{eq:beads_positions} but considering only the $\tilde{\MM}$ lowest-frequency normal modes.
When $\tilde{\MM}=\MM$, the full ring-polymer potential is recovered. On the other hand, when $\tilde{\MM}=1$, ${V_s}$ is only evaluated at the centroid of the ring-polymer. In practice, $\tilde{\MM}$ is an additional convergence parameter that must be checked for each system. As demonstrated in refs.~\cite{fanourgakis2009fast,marsalek2016ab}, the RPC scheme can lead to large gains in performance for some systems. For example, in the case of liquid water modeled via the AMOEBA potential, accurate simulations can be achieved with 
$\MM=32$ for the bonded interactions, and only $\tilde{\MM}\approx5$ for the non-bonded interactions. 
As the non-bonded interactions are much more expensive to compute, this leads to a significant gain in performance. The RPC scheme is of course compatible with the RESPA integrator 
which further reduces the number of required evaluations of the slowly varying forces.
 
\subsubsection{Massively Parallel Implementation}\label{sec:RPMD_parallelization}
\begin{figure}[h!]
    \centering
    \includegraphics[width=1.\linewidth]{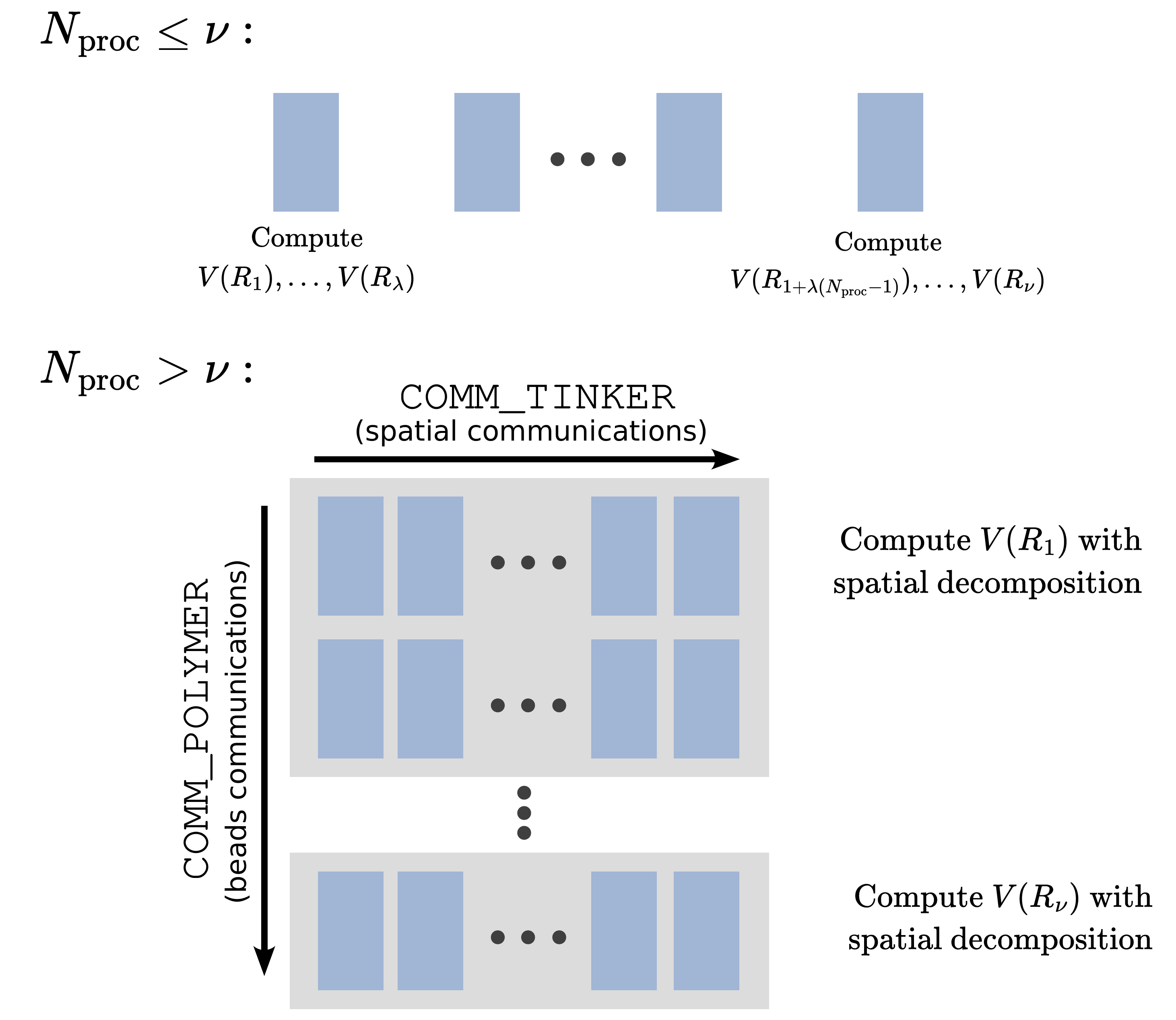}
    \caption{Schematic representation of the parallelization scheme used for the evaluation of the forces in RPMD simulations. The figure distinguishes the two subcases: $N_{proc}\leq \MM$ (top) and $N_{proc}>\MM$ (bottom). In the top figure, we define $\lambda=\MM/N_{proc}$.}
    \label{fig:pi_parallelism}
\end{figure}

The most time-consuming operation in a MD simulation is usually the evaluation of the interatomic forces. It is even more marked in path-integral simulations, where the forces must be evaluated on multiple replicas of the system. However, this evaluation is independent for each bead $x_i^{(\MM)}(\QQ)$ and it is thus efficient to parallelize by assigning the evaluation of the forces on each bead to a different process (or set of processes). When the total number of processes $N_{proc}$ is smaller than the number of beads, each process independently evaluates the forces on a subset of the replicas, as depicted in the top part of figure~\ref{fig:pi_parallelism}. On the other hand, when  $N_{proc}>\MM$, we employ a two-level parallelization scheme that leverages the spatial domain decomposition already implemented in Tinker-HP~\cite{lagardere2018tinker}.
To this aim, the main MPI communicator is split into a grid as schematically shown in the bottom part of Figure.~\ref{fig:pi_parallelism}. The communicator {\verb COMM_POLYMER } (of size $N_{proc}^\ttiny{polymer}$) allows communication between different beads within the same spatial region
, while the communicator {\verb COMM_TINKER } (of size $N_{proc}^\ttiny{spatial}$), that runs horizontally in the figure, allows for communication between different spatial regions at a fixed bead index. 

Once interatomic forces have been evaluated on each bead, they are communicated through {\verb COMM_POLYMER } and projected on the normal modes according to eq.~\eqref{eq:force_normal_modes}.
At this point, the equations of motion for each atom (and each ring-polymer normal mode) can be independently propagated until the next force evaluation is required. This propagation is parallelized by evenly distributing the local atoms among the $N_{proc}^\ttiny{polymer}$ processes of each spatial region. When the centroid of the ring-polymer of an atom changes domains, the information for all its normal modes are transferred to the neighbouring processing units. Neighbour lists are also computed with respect to centroid positions: if the centroids of two atoms are considered neighbours, all the corresponding beads are also considered neighbours. This avoids duplicating neighbour lists for all the beads, thus drastically reducing the associated computational cost and memory requirements.


Note that when using the RPC scheme of section~\ref{subsec:multistep}, the parallelization strategy is defined based on the number $\tilde{\MM}$ of beads in the contracted ring-polymer instead of the full number $\MM$. The evaluation of the slowly varying 
forces (typically the most time-consuming step of the calculation) is then distributed for the contracted ring-polymer with the same parallelization strategy as in Figure~\ref{fig:pi_parallelism}, while for the evaluation of the quickly varying forces, the beads of the full ring-polymer are partitioned using the same spatial decomposition as for the contracted one. 

\subsection{adQTB}
The adQTB implementation uses the standard classical Langevin integrators (BAOAB, BAOAB-RESPA, BAOAB-RESPA1) previously included in Tinker-HP and only replaces the white noise random forces by the adQTB colored noise. 
However, contrary to white noise that can easily be generated on the fly using a standard pseudo-random number generator, colored noise is not memoryless. To generate numerical noise with the adequate memory kernel, the trajectory is split into segments of $N_{seg}$ timesteps (typically $N_{seg}\sim 1000$). At the end of each segment, the adaptation procedure is performed and the colored noise is generated in advance to be used in the next segment. 

\subsubsection{Colored Noise Generation}

\begin{figure}
    \centering
    \includegraphics[width=1.\linewidth]{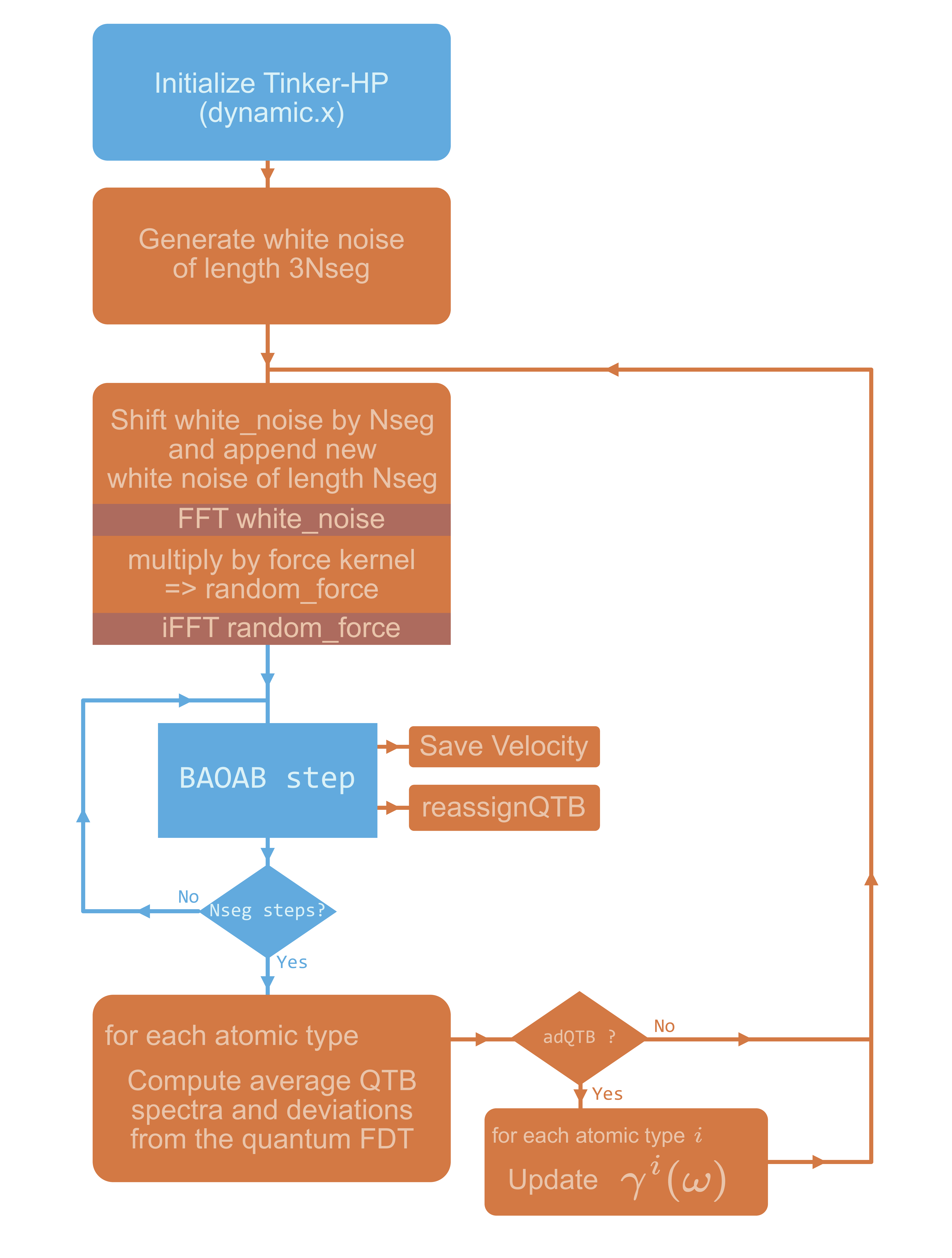}
    \caption{flow chart of a molecular dynamics simulation using the adQTB thermostat.}
    \label{fig:qtbnoise}
\end{figure}

We generate the adQTB colored noise following the \textit{segmented} procedure described in the appendix of ref.~\cite{huppert2022simulation}. In a nutshell, a random force with autocorrelation given by \eqref{eq:Cff_qtb} is computed by performing a convolution between a normalized white noise and the Fourier transform of the square root of \eqref{eq:Cff_qtb} (with corrections for finite timestep~\cite{mangaud2019fluctuation} and non-zero friction~\cite{mauger2021nuclear}). In practice, the convolution is performed in Fourier space (using a standard FFT library) at the beginning of each segment.
Note that in the \textit{segmented} procedure, one needs to store $3N_{seg}$ white noise random numbers for each degree of freedom in order to ensure that the colored noise memory is consistent between segments.
Figure~\ref{fig:qtbnoise} shows a schematic flow chart of the adQTB integration scheme, in which the different steps of the \textit{segmented} noise generation procedure are briefly outlined.

\subsubsection{Computation of the adQTB Spectra and Adaptation Procedure}\label{sec:qtb_spectra_adapt}
As explained in details in refs.~\cite{mangaud2019fluctuation,huppert2022simulation}, the adaptive QTB relies on the quantum fluctuation-dissipation theorem to monitor and compensate ZPE leakage. To that end, we evaluate the deviations from the FDT defined for each degree of freedom $i$ as:
\begin{equation}\label{eq:QTB_spectra}
    \Delta_{FDT,i}(\omega)=m_i C_{v_iv_i}(\omega)\gamma_i(\omega)
        - \Re[C_{v_i F_i}(\omega)]
\end{equation}
The correlation functions are estimated at the end of each segment from the trajectories of $v_i$ and $F_i$: 
\begin{equation}
    \begin{aligned}
    C_{v_iv_i}(\w) &\propto \abs{\tilde{v}_i(\w)}^2\\
    C_{v_iF_i}(\w) &\propto \tilde{v}_i(\w)\tilde{F}^*_i(\w)\\
    \end{aligned}
\end{equation}
where $\tilde{v}_i(\omega)$ and $\tilde{F}_i(\omega)$ are the (discrete) Fourier transforms of the trajectories $v_i(t)$ and $F_i(t)$ over the last segment (the last $N_{seg}$ timesteps, that have thus to be stored in memory). In practice, the values of $\omega$ are discretized consistently with the discrete Fourier transform over the segment, though for simplicity, we will keep the continuous notation for $\omega$ in the following.

In principle, the adjustable parameters of the bath $\gamma_i(\w)$ could be optimized for each degree of freedom in order to cancel each $\Delta_{FDT,i}(\w)$. In practice, we set the same $\gamma_i(\w)$ for all degrees of freedom that share the same atom type number $z$ in Tinker's input parameters, and optimize using the averaged $\overline{\Delta}_{FDT,z}~=~\frac{1}{N_z}\sum_{i\in z}\Delta_{FDT,i}$. This allows to average statistical fluctuations that may affect $\Delta_{FDT,i}$ over all equivalent degrees of freedom, thus improving the convergence of the adaptation procedure.
The implementation provides two adaptation schemes. In the first scheme, denoted as {\verb SIMPLE } and described in details in ref.~\cite{mangaud2019fluctuation}, the coefficients are adapted at the end of each segment according to:
\begin{equation}
    \gamma_z^{(k+1)}(\w)=\gamma_z^{(k)}(\w) - A_{\gamma,z} \overline{\Delta}_{FDT,z}^{(k)}(\w)
\end{equation}
where $\overline{\Delta}_{FDT,z}^{(k)}$ is computed from the $k$-th segment of trajectory and the $\gamma_z^{(k)}$ are the corresponding bath parameters, while the $\gamma_z^{(k+1)}$ are the new parameters to be used in the next segment. The coefficients $A_{\gamma,z}$ allow to adjust the adaptation speed, for each atom type $z$.
The second scheme, denoted as {\verb RATIO }, allows for a faster adaptation of the bath parameters when large numbers of atoms share the same type $z$, while maintaining a controllable level of noise on $\gamma_z$. This adaptation scheme is based on the fact that, if $\gamma_z^\star(\w)$ were the optimal parameters, we would have:
\begin{equation}
     \overline{\Delta}_{FDT,z}^\star(\w)=0 \Leftrightarrow \gamma_z^\star(\w)=
        \frac{\Re[\overline{C}_{vF,z}(\w)]}{m_z \overline{C}_{vv,z}(\w)}
\end{equation}
where $\overline{C}_{vF,z}$ and $\overline{C}_{vv,z}$ are defined similarly as $\overline{\Delta}_{FDT,z}$. Thus, for each type $z$ we define the new parameters as:
\begin{equation}\label{eq:adapt_ratio}
    \gamma_z^{(k+1)}(\w) = \frac{\Re[\overline{C}_{vF,z}^{(k)}(\w)]}{m_z \overline{C}_{vv,z}^{(k)}(\w)}
\end{equation}
The optimal value of $\gamma_z(\w)$ should then be a fixed point of this iterative scheme.
It should be noted that, 
due to numerical noise in the estimators of $\overline{C}_{vF,z}^{(k)}$ and $\overline{C}_{vv,z}^{(k)}$, the estimator of $\gamma_z^\star(\w)$ resulting from the iterative process may be affected by large fluctuations and possibly biased.
To fix this issue, we replace the ratio in eq.~\eqref{eq:adapt_ratio} by a ratio of spectra obtained from a running average with an exponentially decaying window. For example, for $\overline{C}_{vv,z}$:
\begin{multline}
    \expval{\overline{C}_{vv,z}}_{\tau_z}^{(k)}=\expval{\overline{C}_{vv,z}}_{\tau_z}^{(k-1)}e^{-N_{seg}\Delta t/\tau_z}\\
    +\overline{C}_{vv,z}^{(k)}(1-e^{-N_{seg}\Delta t/\tau_z})
\end{multline}
The parameters $\tau_z$ then dictate the adaptation speed and their admissible values critically depend on the level of statistical noise on both spectra, \textit{i.e.} on the number of equivalent degrees of freedom on which they are averaged. As an example, when simulating a large box of liquid water, where all H and all O atoms are equivalent on average, values of $\tau_O$ and $\tau_H$ of the order of 100 fs to 1 ps are sufficient to provide an accurate and fast adaptation (yielding the same parameters as a slow adaptation with the {\verb SIMPLE } method). On the other hand, when simulating an isolated molecule for which the spectra can only be averaged on few atoms, longer adaptation times are required with $\tau_z$ typically of the order of 100~ps. Note that it is possible to combine both adaptation methods, for example by using the {\verb SIMPLE } method to slowly adapt the parameters of a solute molecule while quickly adapting the parameters of the solvent with the {\verb RATIO } scheme. 
~\\

Finally, in order for the random force power spectrum to be well defined, a lower bound $\gamma_\ttiny{min}$ is set on $\gamma_z$ by performing the operation  $\gamma_z^{(k+1)}(\w)~\leftarrow~\max(\gamma_\ttiny{min},~~\gamma_z^{(k+1)}(\w))$ before generating a new segment of colored noise. By default, we set $\gamma_\ttiny{min}=0.01\gamma_0$. As illustrated in ref.~\cite{mangaud2019fluctuation}, this lower bound implies that the ZPE leakage cannot be compensated with an arbitrarily small value of the friction coefficient $\gamma_0$.


\subsubsection{Massively Parallel Implementation}
The parallelization scheme for the adQTB is straightforward as it fully utilizes the spatial decomposition previously implemented in Tinker-HP. The only additional burden compared to classical dynamics is the necessity to keep track of the colored noise for each degree of freedom. Indeed, when an atom is transferred to another cell of the spatial decomposition, its pre-generated colored noise must also be transferred. This corresponds to the "reassignQTB" of Figure \ref{fig:qtbnoise}. 
In order to avoid unnecessary communications, we ensure that colored noise transfer between two processes happens at most once per segment for each atom.

At the end of each segment, the spectra in eq.~\eqref{eq:QTB_spectra} are computed in parallel and averaged for each atom type $z$ on the process of rank zero. The latter then performs the adaptation of the $\gamma_z(\omega)$ as described in section~\ref{sec:qtb_spectra_adapt} and broadcasts the updated parameters to the other processes so that each can then generate the new segment of colored noise for the atoms in its spatial decomposition region. 

\subsection{Extension to GPU Architectures}
Additionally to the massively parallel MPI CPU version, we implemented both methods in the multi-GPU version of Tinker-HP. The critical part of the GPU acceleration, described in ref.~\cite{adjoua2021tinker}, is contained in the calculation of the interatomic forces and did not require any alterations. The GPU port of our methods was done through OpenACC directives in order to offload the generation of the colored noise for adQTB and the integration and normal modes calculations for RPMD onto the device. Much care was taken to suppress unnecessary data transfer between CPU and GPU so that all extra variables (positions and momenta of the normal modes in the case of RPMD and storage of noise and trajectory segments for adQTB) are GPU-resident, \textit{i.e.} are uploaded once on the GPU at the beginning of the simulation and accessed almost exclusively by the GPU. In the case of multi-GPU calculations, direct GPU-to-GPU MPI communications are performed whenever the host architecture allows it.

\subsection{Scaling and Efficiency Tests}
\begin{figure}
    \centering
    \includegraphics[width=0.48\textwidth]{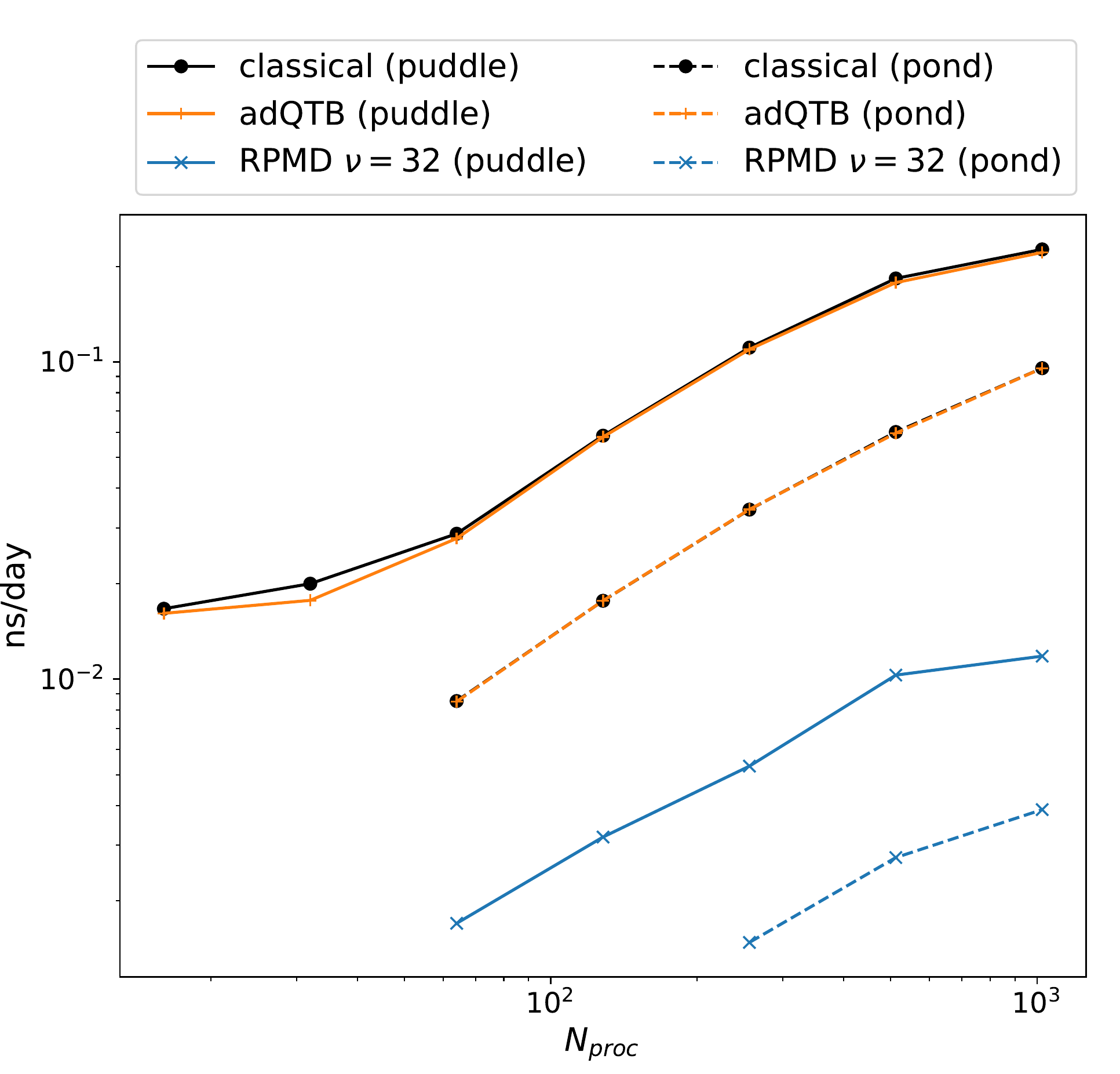}
    \caption{Scaling tests on multi-CPU architecture for the different methods. Performance is indicated by the number of nanoseconds of simulation per walltime day as a function of the number of processes.}
    \label{fig:scaling_CPU}
\end{figure}

We tested the parallelization efficiency on boxes of water of sizes 96000 atoms (puddle) and 288000 atoms (pond). Calculations were performed on the Joliot-Curie cluster located at TGCC and managed by the CEA. We used two of its partitions made of interconnected nodes. Traditional nodes from the first partition are made of 2 AMD Epyc processors with 64 cores each and, clocked at 2.6Ghz. The second partition holds 2 CPUs Intel Cascade Lake of 20 cores each, clocked at 2.1 GHz, and accelerated with 4 GPUs NVIDIA V100 interconnected with NVIDIA NVLink. All simulations use a timestep of 0.2fs, which safely ensures a low integration error for all methods. Figure~\ref{fig:scaling_CPU} shows performance (measured in nanoseconds of simulation per day of computation) as a function of the number of CPUs in a log-log scale for both system sizes. We first notice that the performance of the adQTB are almost identical to that of classical MD, confirming that the colored noise generation and the adaptation scheme only make a small contribution to the computation time for these moderately large systems. The raw performance of the RPMD is of course lower than that of classical MD (due to the 32 replicas used for the simulation) but the scaling with the number of processes is similar. Note that for these simulations, there are more processes than RPMD replicas so that the two-level bead/spatial parallelization described in section~\ref{sec:RPMD_parallelization} is fully used.

\begin{figure}
    \centering
    \includegraphics[width=0.5\textwidth]{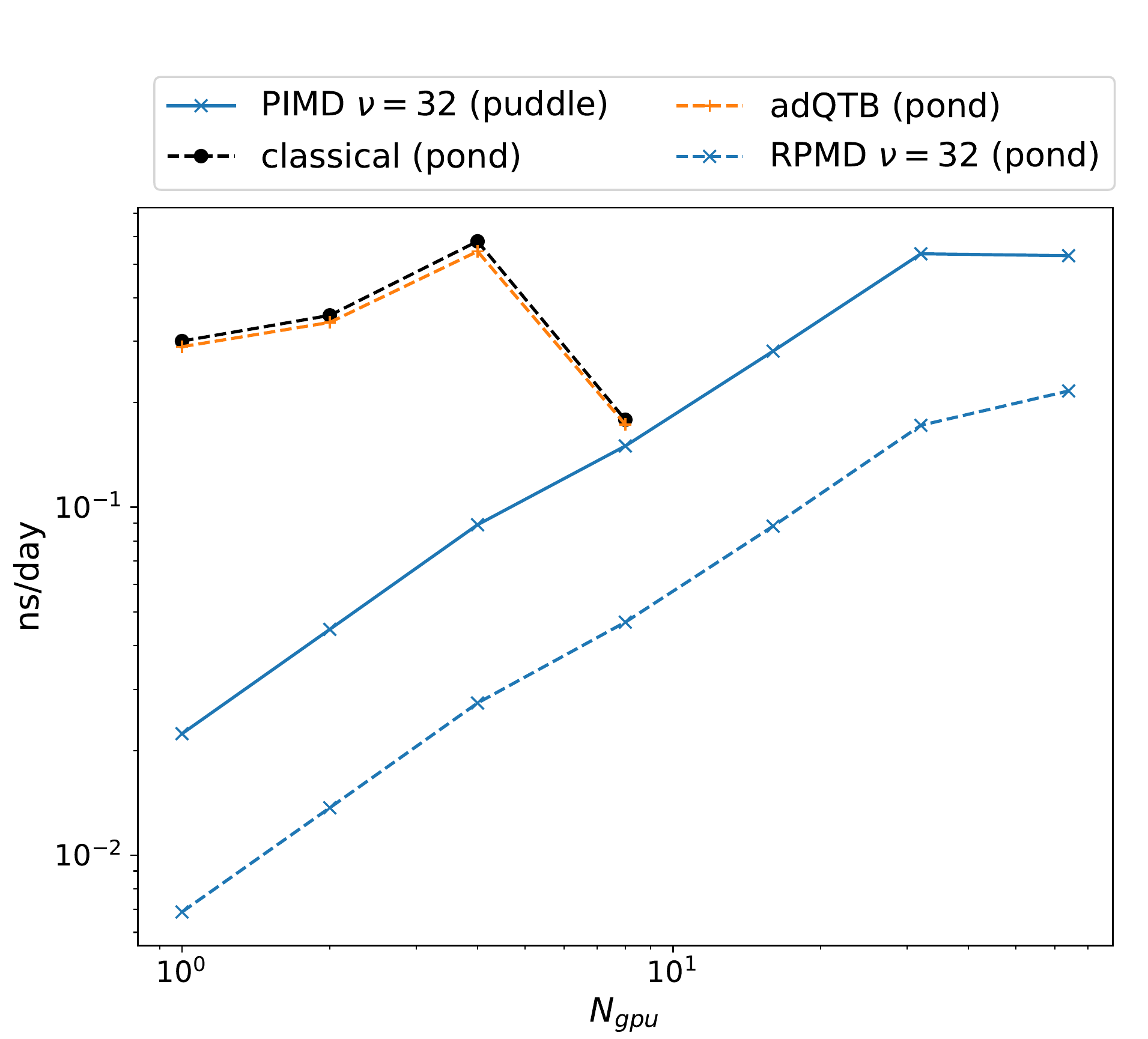}
    \caption{Scaling tests on multi-GPU architecture for the different methods. Performance is indicated by the number of nanoseconds of simulation per walltime day as a function of the number of processes. Nodes are composed of four interconnected V100 GPUs so that when using more than four GPUs, out-of-node communications are required, causing a drop in the efficiency.}
    \label{fig:scaling_GPU}
\end{figure}

Figure~\ref{fig:scaling_GPU} shows the performance of the same methods on multi-GPU architecture. Again, we obtain very similar performance in classical MD and in adQTB and a very significant performance increase compared to the CPU architecture. The drop in performance when going from four to eight GPUs is due to inefficiencies in the out-of-node communications (nodes at TGCC are composed of four interconnected V100 GPUs) which have a critical impact for the spatial decomposition parallelization scheme. On the other hand, multi-node parallelization in RPMD remains very efficient as long as the number of GPUs is smaller than the number of replicas since the interatomic forces are then evaluated in parallel with very few communications compared to a purely spatial decomposition.

\section{Perspective 1: Inclusion of Quantum Nuclear effects in Machine-learning Potentials Simulations using the Deep-HP Platform}\label{sec:deephp}
Our platform for nuclear quantum effects is fully compatible with the recently developed Deep-HP module~\cite{inizan2022scalable} of Tinker-HP that enables the use of machine-learning potentials (MLPs), such as ANI~\cite{smith2017ani,devereux2020extending} or DeePMD~\cite{zhang2018deep,ko2019isotope}, to perform molecular dynamics simulations. It also enables hybrid machine-learning/physical force field calculations in a QM/MM-like embedding framework. MLPs in principle require the explicit inclusion of nuclear quantum effects to achieve their best accuracy on thermodynamical properties since they usually are fitted solely on \textit{ab initio} data. It is thus of the utmost importance for future developments of MLPs to be able to efficiently perform quantum MD in order to assess their accuracy. Since the computational cost of MLPs, as of today, is about an order of magnitude greater than that of polarizable force fields such as AMOEBA, coupling them with path integrals requires a lot of computational resources (especially since integration tricks such as multi-timestepping or RPC cannot usually be used for these potentials). The adQTB, on the other hand, provides a much cheaper alternative that allows to quickly compute thermodynamical properties with good accuracy, as demonstrated in previous litterature~\cite{mauger2021nuclear} and as we will further show in section~\ref{sec:applications}. 

\begin{figure}
    \centering
    \includegraphics[width=0.45\textwidth]{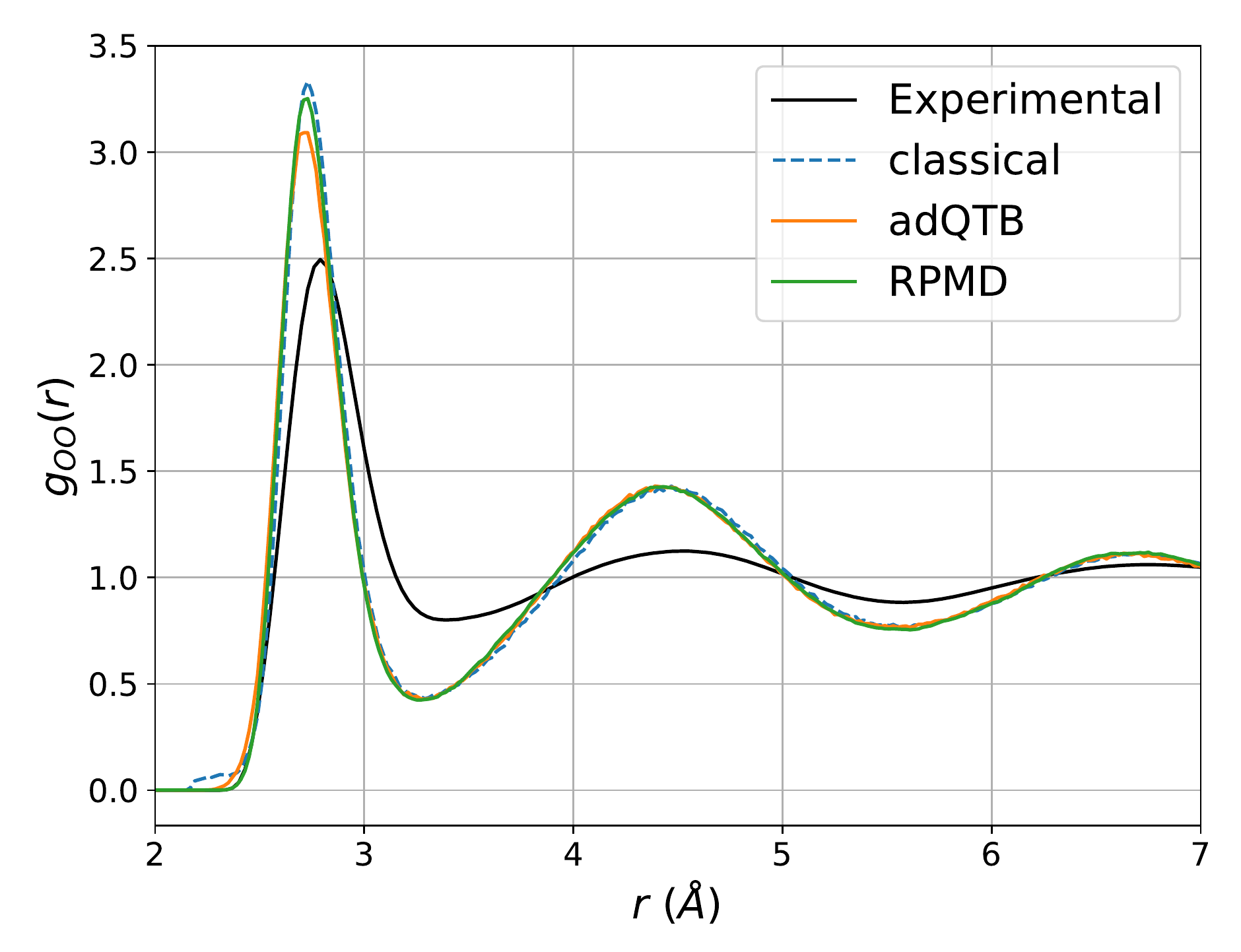}
    \caption{Oxygen-Oxygen radial distribution function of water at 300K computed using the DeePMD ML force field and simulated with classical dynamics (dashed), adQTB (solid orange) and RPMD (solid green). Experimental results from ref.~\cite{soper2013radial}. }
    \label{fig:gOO_deepmd}
\end{figure}

In this section, we show the compatibility of Quantum-HP and Deep-HP by computing radial distribution functions (RDFs) of liquid water using the DeePMD potential. We performed 500 ps of NVT simulation (at experimental density) for a cubic box of 1000 water molecules for both classical and adQTB MD, and a smaller box of 216 molecules for PIMD (with $\MM=32$ beads). Figure~\ref{fig:gOO_deepmd} shows the Oxygen-Oxygen RDF of liquid water simulated with adQTB, RPMD and classical MD compared with experimental data from ref.~\cite{soper2013radial}. The DeePMD model was trained on path integral \textit{ab-initio} molecular dynamics (PI-AIMD) trajectories, at the PBE0-TS level (refs~\cite{hsinYu2019isotope,zhang2021phase}): (1) 100000 snapshots of PI-AIMD liquid water (192 atoms) at 1 bar and 300K (2) 20000 snapshots of PI-AIMD ice phase Ih (288 atoms) at 1 bar and 273K (3) 10000 snapshots of classical AIMD ice phase Ih at 1 bar and 330K and (4) 10000 snapshots of classical AIMD ice phase Ih at 2130 bar and 238K. We used 10\% of the data as validation set. The DeePMD model was trained using the DeePMD-kit package~\cite{zhang2018deep}. The DeePMD model architecture is composed of a (25, 50, 100) embedding net with a 18 neuron-size embedding sub-matrix, and a (240, 240, 120, 60, 30, 10) fitting net. The cut-off radius was set to 6~\AA~with a smoothing cutoff of 0.5~\AA~and a two-body embedding descriptor. The final model is trained with $1.2\times10^{7}$ Adam steps. With this training setup, the dynamics was stable and the radial distribution function is in acceptable agreement with experimental results. We note that NQEs appear to be nearly negligible on figure~\ref{fig:gOO_deepmd}. This can be explained by an almost perfect compensation between competing NQEs
~\cite{habershon2009competing,ceriotti2016nuclear,li2011quantum}: the zero-point energies of bending and stretching modes tend to affect the hydrogen bond strength in opposite ways, but the net effect on the structure of the liquid is very small for this particular water model. This net effect is indeed strongly model-dependent and can sometimes attenuate the structure 
of the liquid as in Q-AMOEBA~\cite{mauger2022improving} or 
reinforce it as in  MB-Pol~\cite{li2022static}. NQEs are more noticeable on the O-H and H-H RDFs (provided in Supplementary Information), especially for peaks corresponding to intramolecular distances which display a strong broadening due to large zero-point energy effects. Since the use of neural networks will be the focus of several of our further works, we limit ourselves here concerning the tests but we can already conclude that the Quantum-HP platform can now be used together with the Deep-HP module to efficiently fuel Deep Neural Networks simulations including explicit NQEs.  

\section{Perspective 2: Inclusion of Quantum Nuclear Effects in Polarizable simulations: Application to Hydration Free Energies of Small Organic Molecules}\label{sec:applications}
In this last section, we illustrate the capabilities of the platform by computing hydration free energies (HFE) of small organic molecules using the adQTB method. We demonstrate state-of-the-art accuracy with the recently developed Q-AMOEBA water potential for the solvent and Poltype parametrization of the solutes~\cite{walker2022automation} on a benchmark of forty of the most common organic molecules~\cite{ren2011polarizable,wu2012automation}.\\

Let us first focus on the estimation of free energy differences within quantum simulations. For this study, we used the Bennett Acceptance Ratio (BAR) method~\cite{bennett1976efficient} that can readily be generalized to the path-integrals formalism. Let us denote $V_A$ and $V_B$ the potential energies of two thermodynamical states. The free energy difference between the two states is defined as $\Delta F_{AB}=\beta^{-1}\ln(Z_A/Z_B)\approx\beta^{-1}\ln(Z_{A,\MM}/Z_{B,\MM})$ with $Z_A$ and $Z_B$ the quantum partition functions of states $A$ and $B$ and their respective path-integral counterparts $Z_{A,\MM}$ and $Z_{B,\MM}$ (note that we recover the equality in the $\MM\to\infty$ limit). The Path Integral Bennett Acceptance Ratio (PI-BAR) estimator of the free energy difference is then given by:
\begin{align}
    \Delta F_{AB} &= C + \beta^{-1}\ln 
  \frac{\expval{ f_\beta(U_{A,\MM} - U_{B,\MM} +C)}_{B,\MM}}
       {\expval{f_\beta(U_{B,\MM} - U_{A,\MM} -C)}_{A,\MM}} \label{eq:PIBAR1}\\
    C&=\Delta F_{AB} + \beta^{-1}\ln(n_B/n_A)\label{eq:PIBAR2}
\end{align}
with $f_\beta(x)=(1+\exp(\beta x))^{-1}$, $U_{A,\MM}$ and $U_{B,\MM}$ defined as in eq.~\eqref{eq:PI_potential} and $n_B$ and $n_A$ the sample sizes used to estimate the corresponding averages. Note that \eqref{eq:PIBAR1} and \eqref{eq:PIBAR2} form a self-consistent set of equations that is solved iteratively.

Although eq.~\eqref{eq:PIBAR1} is the minimal expected variance estimator for $\Delta F_{AB}$~\cite{bennett1976efficient}, its accuracy still relies on a somewhat large overlap between the probability distributions of states A and B. Thus, direct estimation of hydration free energies (defining state A as the molecule in solution and state B as the gas phase) is in general impossible~\cite{wu2005phase1}. In line with standard procedures~\cite{wu2005phase2,shi2011multipole}, we compute the hydration free energy as a sum of free energy differences between neighbouring states in a thermodynamical path that progressively decouples the solute from the solvent. First, the electrostatic and polarization interactions between the solute and the solvent are turned off by progressively scaling down the permanent multipoles and polarizabilities of the solute. Then, the van der Waals interactions between the solute and the solvent are scaled down to zero (while using a soft-core potential~\cite{steinbrecher2011soft,lagardere2018tinker}). To recover the hydration free energies, the solute is then "recharged" in the gas phase (\textit{i.e.} the intramolecular electrostatic interactions are turned back on progressively).\\

To check the consistency of the methods and the accuracy of the Q-AMOEBA water potential, we first computed the solvation free energy of a Q-AMOEBA water molecule in Q-AMOEBA water. We used a progressive decoupling with 20 thermodynamic states (the precise decoupling schedule that we used is provided in Supporting Information) that were all simulated using a BAOAB-RESPA integrator with an inner timestep of 0.2 fs and an outer timestep of 2 fs in the NVT ensemble at 300K and experimental density. For each thermodynamical state, we thermalize the system for 1 ns and accumulate statistics for 3 ns. The PI-BAR method yields a free energy difference of -5.70$\pm$0.05 kcal/mol while the classical BAR value is -6.44$\pm$0.04 kcal/mol, demonstrating the strong influence of nuclear quantum effects on the HFE. We note that, for the original AMOEBA force field in classical MD, the HFE was previously reported at -5.86$\pm$0.19~\cite{ponder2010current}. 
Thus, as could be expected, the Q-AMOEBA results with explicit NQEs are close to that of classical simulations with AMOEBA, which was fitted in such a way that it implicitly includes NQEs. On the other hand, the experimental HFE for water was measured at -6.32 kcal/mol. 
The underestimation of the absolute value of the HFE by Q-AMOEBA is consistent with previous results reported for the enthalpy of vaporization (underestimated by approximately 1 kcal/mol~\cite{mauger2022improving}) and the general interpretation that Q-AMOEBA slightly underestimates the strength of hydrogen bonds. We also performed path integrals simulations with a two-stage contraction scheme (with long-range forces and polarization estimated on the centroid only, non-bonded short-range forces evaluated on 12 beads and bonded forces on the whole 32 beads polymer) and obtained a HFE of -5.84$\pm$0.05 kcal/mol, in good agreement with the complete 32 beads calculation, while saving a factor $\sim $6 on computation time.

\begin{figure}
    \includegraphics[width=0.45\textwidth]{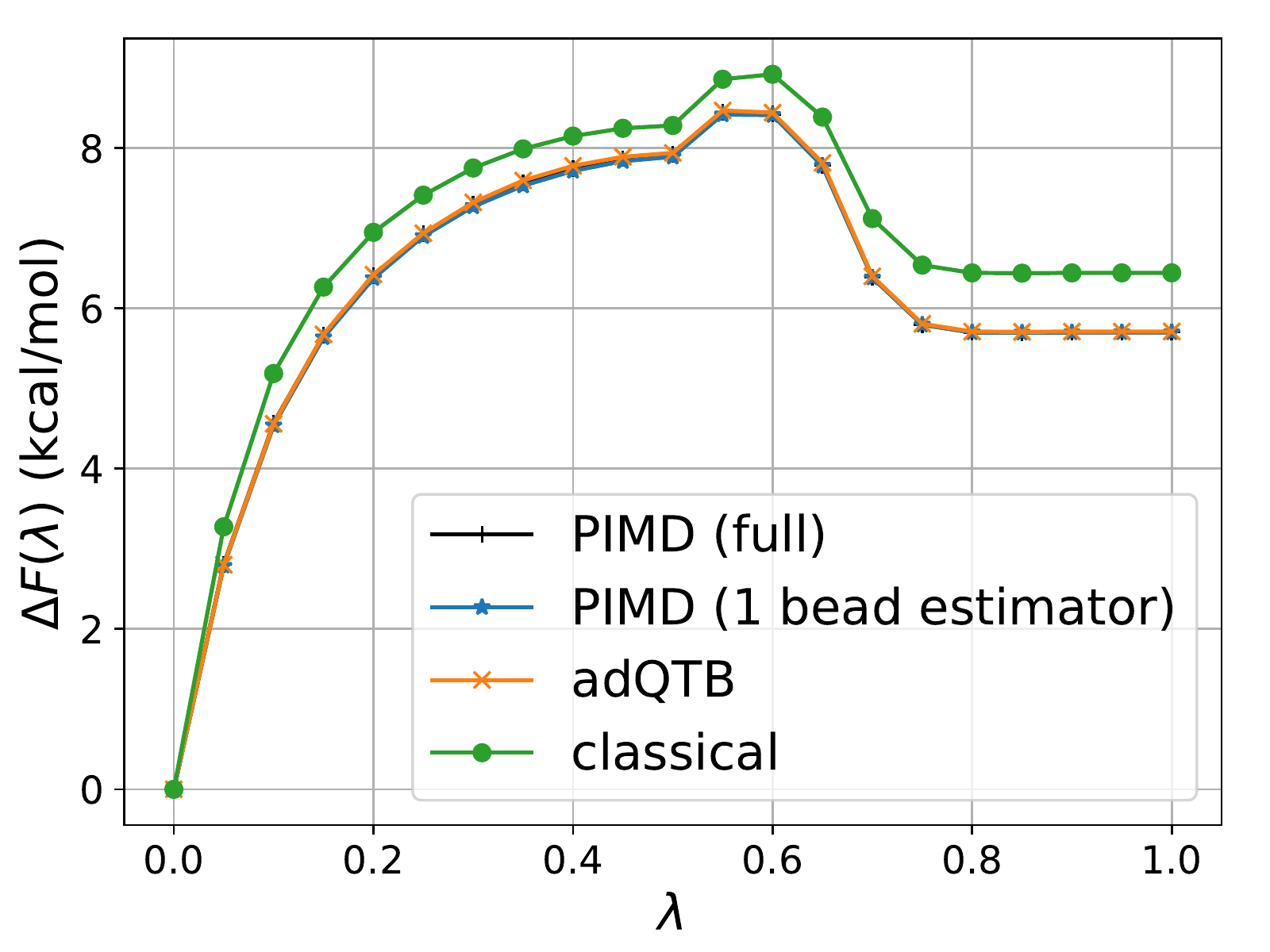}
    \caption{Q-AMOEBA potential of mean force along the vaporization thermodynamical path. $\lambda=0$ corresponds to the fully solvated molecule and $\lambda=1$ corresponds to non-interacting solute/solvent.}
    \label{fig:water_PMF}
\end{figure}

While an unbiased estimator of free energy differences can analytically be derived from the path-integral partition function (equation~\eqref{eq:PIBAR1}), this is not the case for the adaptive QTB. Previous work, however, showed that the probability distribution sampled by the adQTB is usually very close to that of a single bead of the ring-polymer (\textit{i.e.} the correct quantum distribution) such that the estimation of configurational averages with the adQTB is in general accurate. Equation \eqref{eq:PIBAR1}, however, is peculiar as it involves the average value of a non-linear function of the bead-averaged potentials $U_{A,\MM}$, $U_{B,\MM}$. In principle, it could therefore be affected by instantaneous correlations between the beads that the adQTB cannot capture. In practice, however, one can show (see appendix~\ref{app:1PIBAR}) that 
the bias induced by
replacing the bead-averaged potential in equation \eqref{eq:PIBAR1} by the value of the potential on a single bead
is of order at least two 
in the potential energy difference $V_A-V_B$ (\textit{i.e.} negligible when the decoupling is sufficiently gradual). Indeed, we verified numerically that in the case of Q-AMOEBA water the single-bead HFE is statistically indistinguishable from the unbiased estimator (see Figure~\ref{fig:water_PMF}). Thus correlations between beads seem only to play a minor role in the free energy estimation and, in turn, the adQTB should provide accurate free energy differences using the standard BAR estimator. Indeed, the HFE for water computed using the adQTB is -5.71$\pm$0.04 kcal/mol, in excellent agreement with the full path-integrals calculation. Figure \ref{fig:water_PMF} shows the potential of mean force along the thermodynamical path and demonstrates that the accuracy of the adQTB (and of the single bead estimator) is not due to error compensations along the path and that the free energy difference at each window is indeed accurately estimated.

\begin{figure}
    \centering
    \includegraphics[width=0.5\textwidth]{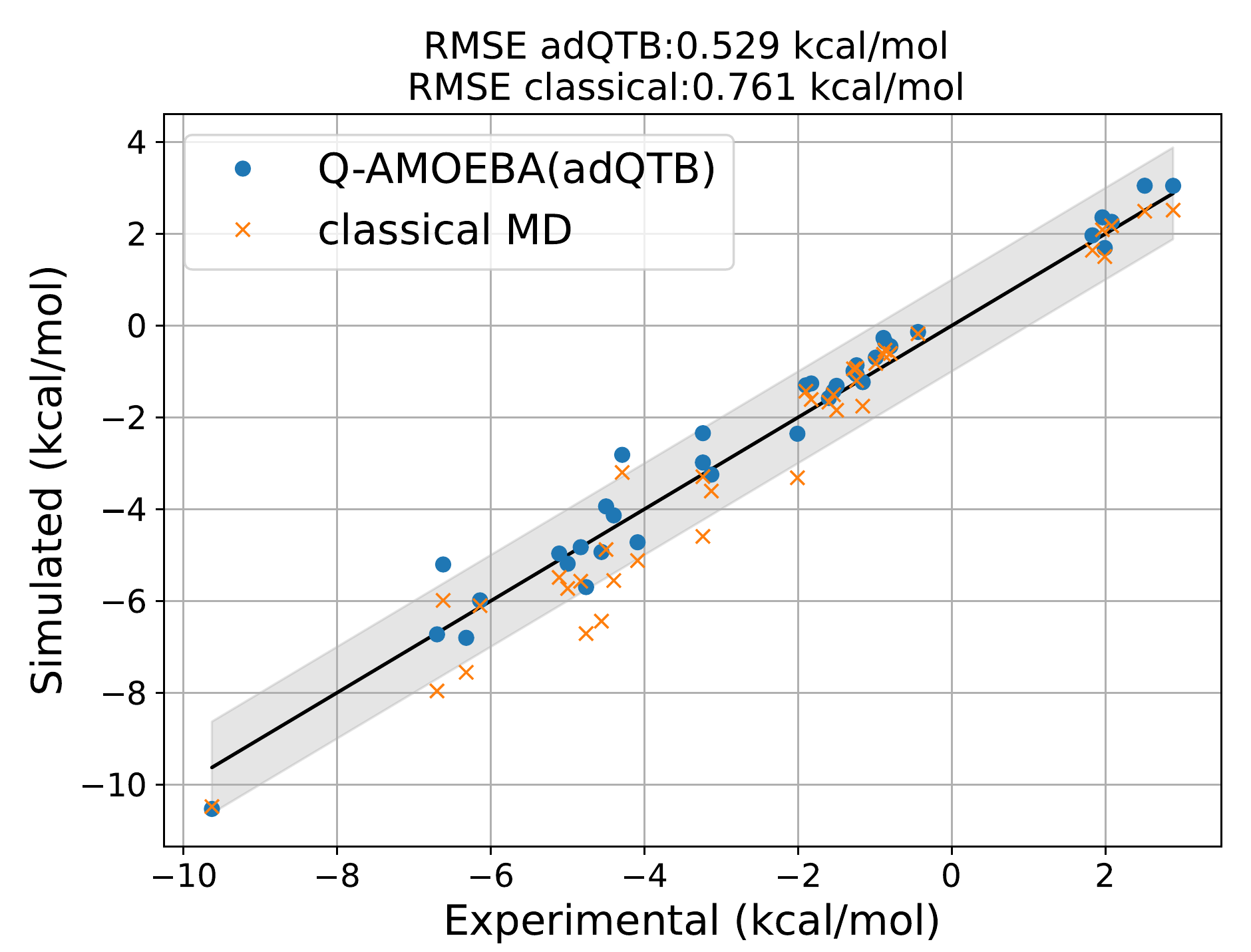}
    \caption{Q-AMOEBA Hydration free energies of small organic molecules simulated with adQTB and classical MD compared to experimental values (experimental data and molecules parametrization from refs.~\cite{walker2022automation,ren2011polarizable})}
    \label{fig:HFE_molecules+expt}
\end{figure}

We then proceeded to compute the hydration free energies for a benchmark of approximately forty small organic molecules. All simulations were performed using the same setup as for the calculations on water. We used Q-AMOEBA to model the solvent and the solutes were parametrized using the Poltype2 software~\cite{walker2022automation}, except for methylamine and dimethylamine for which AMOEBA09 parameters~\cite{ren2011polarizable} were used. Figure~\ref{fig:HFE_molecules+expt} shows a scatter plot of the adQTB and classical HFE against experimental values. We clearly see as a systematic trend that nuclear quantum effects tend to hinder solvation, which is likely due to a weakening of hydrogen bonding between the solute and solvent when including NQEs. 
Including NQEs also brings the results closer to the experimental values, with a correlation coefficient $r^2$ of 0.97 and a root mean square error (RMSE) over the dataset of 0.52 kcal/mol using the adQTB acompared to $r^2=$0.93 and a RMSE of 0.76 kcal/mol when neglecting NQEs. We also note that our results are in slightly better agreement with the experimental data than those previously reported 
over the same solutes dataset, but without explicit inclusion of NQEs  (RMSE of 0.58 kcal/mol~\cite{walker2022automation} using the original AMOEBA parametrization for the solvent that implicitly includes NQEs).  Importantly, this improvement has been obtained without fitting the parameters of the force field on the experimental HFEs, thus reinforcing the idea that explicitly taking into account NQEs allows for the development of more transferable models.

\section{Conclusion}\label{sec:conclusion}
We introduced a new platform inside Tinker-HP that enables the explicit inclusion of nuclear quantum effects (NQEs) in molecular dynamics (MD) simulations.
The platform, denoted Quantum-HP, implements two methods for quantum MD: ring-polymer MD and adaptive quantum thermal bath MD. While the former provides exact reference results at a relatively high computational cost, the latter was previously shown to give a reliable approximation of NQEs~\cite{mauger2021nuclear} and was the method of choice for the development of the new Q-AMOEBA force field~\cite{mauger2022improving}.

The Quantum-HP platform is massively parallel and supports multi-GPU architectures. We have shown that the cost and scaling of the adQTB method is almost identical to that of classical MD and that path integrals, although more expensive, display excellent scaling up to thousands of CPUs and hundreds of GPUs thanks to a two-level parallelization scheme. This makes path-integral MD on Tinker-HP a good candidate to be able to harness the computational power of exascale machines for simulations with new generation models.

We demonstrated the applicability of our platform for the computation of the hydration free energy of small ligands. In these simulations, the solvent was modeled using the newly introduced Q-AMOEBA~\cite{mauger2022improving} potential while the ligands were parametrized using the Poltype2~\cite{walker2022automation} software. We showed that the explicit inclusion of NQEs improves the accuracy of such free energies so that future models should be designed with this knowledge, while the additional cost is still affordable when using the adQTB. 

The efficiency and massive parallelization capabilities of the newly introduced platform now allows the inclusion of explicit nuclear quantum effects in very large systems. This should be particularly relevant for simulations in extreme conditions of pressure or temperature where NQEs can be massive~\cite{markland2018nuclear, benoit1998tunnelling,schaack2020quantum, witt2010path} or to investigate disorder effects in solids for which NQEs can be determinant and large supercells required~\cite{geneste2013low, Bronstein_PRB2016, fallacara2021thermal}. Importantly, being able to simulate quantum nuclei enables the study of isotope effects that are simply not reachable using classical MD~\cite{ceriotti2013efficient,cheng2014direct}. Finally, it opens up the possibility of investigating quantitatively the importance of NQEs in biological processes~\cite{fang2016inverse,law2018importance} and the subtleties of hydrogen-bonded systems~\cite{li2011quantum}. While the methods for quantum MD are now readily available in Tinker-HP, it will be necessary to re-parametrize some of the force fields to avoid double counting of implicit and explicit NQEs, as was shown in the case of Q-AMOEBA for water~\cite{mauger2022improving}. The parametrization of Q-AMOEBA for ions, organic molecules and biomolecules will thus be at the forefront of near future developments. In addition, it will also enable explicit NQEs simulations with advanced polarizable potentials. Models natively designed to reproduce the Born-Oppenheimer surface such as SIBFA \cite{Gresh2007, naseem2022development} should directly be applicable whereas a re-parametrization of aproaches such as AMOEBA+~\cite{liu2019amoebaplus} or HIPPO~\cite{rackers2021polarizable} will be necessary. Neural networks such as ANI\cite{smith2017ani}, DeePMD\cite{zhang2018deep}, Physnet \cite{PhysNet} (and others) can also be used directly with Quantum-HP. Therefore, with the improvements in computing power and the availability of the methods, we expect that explicit NQEs will soon be an integral part of the standard workflow for both the force field developer and the MD practitioner. Furthermore, the platform will be naturally extended to methods that rely on the simultaneous simulation of multiple replicas of one system such as replica-exchange\cite{sugita1999replica}, adaptive sampling \cite{inizan2021high} or adaptive bias methods using multiple walkers\cite{minoukadeh2010potential}.


\section*{Acknowledgements}
This work has received funding from the European Research Council (ERC) under the European Union’s Horizon 2020 research and innovation program (grant agreement No 810367), project EMC2 (JPP). Computations have been performed at GENCI on the Jean-Zay machine (IDRIS, Orsay, France) and on the Joliot-Curie cluster (TGCC, Bruyères le Châtel, France) on grant no. A0070707671.
\\ \\
* thomas.ple@sorbonne-universite.fr, \\
* louis.lagardere@sorbonne-universite.fr,\\
* jean-philip.piquemal@sorbonne-universite.fr,\\

\section*{Supporting Information Available}
The supporting information contains the decoupling schedules usef for estimating the hydration free energies presented in the main text, the values and standard errors of the hydration free energies for the dataset of small organic molecules and the three radial distribution functions of liquid water computed using the DeePMD neural network potential.

\bibliography{biblio} 

\clearpage
\appendix
\section{Single-bead PI-BAR Estimator}\label{app:1PIBAR}
We define the single-bead PI-BAR estimator as:
\begin{equation}
    \Delta F_{AB}^\ttiny{(1 bead)} = C + \beta^{-1}\ln 
  \frac{\expval{ f_\beta(V_A - V_B +C)}_{B,\MM}}
       {\expval{f_\beta(V_B - V_A -C)}_{A,\MM}} \label{eq:1PIBAR1}
\end{equation}
where $V_A=V_A(x_0^{(\MM)}(\QQ))$ is the potential energy of state A estimated at the position of a single bead of the ring-polymer (the choice of the bead index is arbitrary thanks to the cyclic permutation invariance of the ring polymer). In the following, we will show that this estimator is biased with respect to eq.~\eqref{eq:PIBAR1} only to second order (at least) in the difference $\Delta V=V_A - V_B$. For this, let us denote $r_{AB}$ the ratio of average values in eq.~\eqref{eq:1PIBAR1} and write it in terms of explicit integrals over the corresponding distributions:
\begin{equation}
    r_{AB}=
    \frac{Z_{A,\MM}}{Z_{B,\MM}}\frac{
       \int \dd \QQ~f_\beta(\Delta V +C)~e^{-\beta(U_{B,\MM}+K)}
    }{
       \int \dd \QQ~f_\beta(-\Delta V -C)~e^{-\beta (U_{A,\MM}+ K)}
    }
\end{equation}
with $K=\sum_{n>0} \frac{1}{2}\omega_n^2Q_n^T M Q_n$ the harmonic potential of the ring-polymer.
We now use the property of the Fermi function $f_\beta(x)=f_\beta(-x)e^{-\beta x}$ to obtain:
\begin{multline}
    r_{AB}=
    \frac{Z_{A,\MM}e^{-\beta C}}{Z_{B,\MM}}\\
    \times\frac{
       \int \dd \QQ~f_\beta(-\Delta V -C)~e^{-\beta (U_{A,\MM}+K)} e^{-\beta(\Delta V - \Delta U_\MM)}
    }{
       \int \dd \QQ~f_\beta(-\Delta V -C)~e^{-\beta (U_{A,\MM}+K)}
    }
\end{multline}
where $\Delta U_\MM=U_{A,\MM}-U_{B,\MM}=\sum_{i=0}^{\MM}\Delta V(x_i^{(\MM)}(\QQ))/\MM$ the average potential energy difference over the beads of the ring polymer. Expanding the term $e^{-\beta(\Delta V - \Delta U_\MM)}$ in the numerator then gives:
\begin{multline}
    r_{AB}= \frac{Z_{A,\MM}e^{-\beta C}}{Z_{B,\MM}}\Bigg(1 -\beta
      \frac{ \expval{
        f_\beta(-\Delta V -C)(\Delta V -\Delta U_\MM)
      }_{A,\MM}}{\expval{
        f_\beta(-\Delta V -C)
      }_{A,\MM}}\\
      +\order{\Delta V^2}
    \Bigg)
\end{multline}
The only contribution to the first order comes from the order zero in the 
 expansion of the Fermi function which is $f_\beta(-\Delta V~-~C)~=~f_\beta(-C)~+~\order{\Delta V }$. Since $C$ is a constant in the integration over $\QQ$, we obtain:
\begin{equation}
    r_{AB}=\frac{Z_{A,\MM}e^{-\beta C}}{Z_{B,\MM}}\Bigg(1 -\beta\qty(\expval{\Delta V}_{A,\MM} -\expval{\Delta U_\MM}_{A,\MM} )+ \order{\Delta V^2}\Bigg)
\end{equation}
Due to the cyclic permutation invariance of the ring-polymer, we have $\expval{\Delta V}_{A,\MM}=\expval{\Delta U_\MM}_{A,\MM}$ so that the first order cancels out. Plugging back $r_{AB}$ in the single-bead estimator \eqref{eq:1PIBAR1}, we then see that $\Delta F_{AB}^\ttiny{(1 bead)}$ is unbiased at least up to second order in $\Delta V$.

\end{document}


\title[]{Supporting Information for: Routine Molecular Dynamics Simulations Including Nuclear Quantum Effects: from Force Fields to Machine Learning Potentials}
\author{Thomas Plé*}
\affiliation{Sorbonne Université, LCT, UMR 7616 CNRS, F-75005, Paris, France}
\author{Nastasia Mauger}
\affiliation{Sorbonne Université, LCT, UMR 7616 CNRS, F-75005, Paris, France}
\author{Olivier Adjoua}
\affiliation{Sorbonne Université, LCT, UMR 7616 CNRS, F-75005, Paris, France}
\author{Théo Jaffrelot-Inizan}
\affiliation{Sorbonne Université, LCT, UMR 7616 CNRS, F-75005, Paris, France}
\author{Louis Lagardère*}
\affiliation{Sorbonne Université, LCT, UMR 7616 CNRS, F-75005, Paris, France}
\author{Simon Huppert}
\affiliation{Institut des Nanosciences de Paris (INSP), CNRS UMR 7588 and Sorbonne Université, F-75005, Paris, France}
\author{Jean-Philip Piquemal*}
\affiliation{Sorbonne Université, LCT, UMR 7616 CNRS, F-75005, Paris, France}

\maketitle

\section{Hydration free energies of small organic molecules}

\begin{table}[h!]
  \begin{tabular}{|c|c|c|c|c|c|c|c|c|c|c|c|c|c|c|c|c|c|c|c|c|c|}
     \hline
      State no. & 1 & 2 & 3 & 4 & 5 & 6 & 7 & 8 & 9 & 10 & 11 & 12 & 13 & 14 & 15 & 16 & 17 & 18 & 19 & 20 & 21  \\
     \hline
     $\lambda_\ttiny{elec}$ & 1.0 & 0.9 & 0.8 & 0.7 & 0.6 & 0.5 & 0.4 & 0.3 & 0.2 & 0.1 & 0.0 & 0.0 & 0.0 & 0.0 & 0.0 & 0.0 & 0.0 & 0.0 & 0.0 & 0.0 & 0.0  \\
     \hline
     $\lambda_\ttiny{VdW}$ & 1.0 & 1.0 & 1.0 & 1.0 & 1.0 & 1.0 & 1.0 & 1.0 & 1.0 & 1.0 & 1.0 & 0.9 & 0.8 & 0.7 & 0.6 & 0.5 & 0.4 & 0.3 & 0.2 & 0.1 & 0.0  \\
     \hline
  \end{tabular}
  \caption{Decoupling schedule used in the hydration free energy calculations. $\lambda_\ttiny{elec}$ controls the decoupling of electrostatic interactions. $\lambda_\ttiny{VdW}$ controls the decoupling of Van der Waals interactions.}
  \label{tab:decoupling_schedule}
\end{table}

\begin{table}[h!]
  \begin{tabular}{|c|c|c|c|c|c|c|c|c|c|c|c|}
     \hline
      State no. & 1 & 2 & 3 & 4 & 5 & 6 & 7 & 8 & 9 & 10 & 11  \\
     \hline
     $\lambda_\ttiny{elec}$ & 0.0 & 0.1 & 0.2 & 0.3 & 0.4 & 0.5 & 0.6 & 0.7 & 0.8 & 0.9 & 1.0  \\
     \hline
  \end{tabular}
  \caption{Gas-phase recharge schedule used in the hydration free energy calculations.}
  \label{tab:recharge_schedule}
\end{table}
\clearpage
\begin{table}[h!]
  \begin{tabular}{|c|c|c|c|}
\hline
 &  Expt. & adQTB & classical\\
\hline
22-dimethylbutane         &         2.51 &         3.05 (0.04) &         2.49 (0.07)\\
aceticacid                &        -6.70 &        -6.73 (0.05) &        -7.96 (0.05)\\
benzene                   &        -0.87 &        -0.35 (0.06) &        -0.54 (0.05)\\
diethylsulfide            &        -1.60 &        -1.58 (0.07) &        -1.67 (0.07)\\
dimethylamine             &        -4.29 &        -2.82 (0.04) &        -3.20 (0.04)\\
dimethyldisulfide         &        -1.83 &        -1.26 (0.06) &        -1.61 (0.06)\\
dimethylsulfide           &        -1.54 &        -1.43 (0.05) &        -1.51 (0.05)\\
di-n-butylamine           &        -3.24 &        -2.98 (0.05) &        -4.59 (0.10)\\
di-n-propylether          &        -1.16 &        -1.23 (0.08) &        -1.76 (0.09)\\
di-n-propylsulfide        &        -1.28 &        -0.99 (0.08) &        -0.94 (0.09)\\
ethane                    &         1.83 &         1.97 (0.04) &         1.64 (0.04)\\
ethanol                   &        -5.00 &        -5.19 (0.05) &        -5.73 (0.04)\\
ethylamine                &        -4.50 &        -3.94 (0.05) &        -4.88 (0.04)\\
ethylbenzene              &        -0.80 &        -0.45 (0.07) &        -0.61 (0.08)\\
hydrogensulfide           &        -0.44 &        -0.14 (0.03) &        -0.18 (0.03)\\
imidazole                 &        -9.63 &       -10.53 (0.05) &       -10.48 (0.05)\\
isopropanol               &        -4.76 &        -5.70 (0.05) &        -6.71 (0.05)\\
\hline
  \end{tabular}
  \caption{Experimental, adQTB and classical values of the HFE (in kcal/mol)} for the dataset of organic molecules (alphabetical order, from A to L). Estimated standard errors are given in parenthesis.
  \label{fig:hfe_values}
\end{table}
\clearpage
\begin{table}[h!]
  \begin{tabular}{|c|c|c|c|}
\hline
 &  Experiment & adQTB & classical\\
\hline
methane                   &         1.99 &         1.69 (0.03) &         1.50 (0.03)\\
methanethiol              &        -1.24 &        -0.87 (0.04) &        -0.93 (0.04)\\
methanol                  &        -5.11 &        -4.97 (0.04) &        -5.49 (0.04)\\
methylacetate             &        -3.13 &        -3.25 (0.06) &        -3.61 (0.05)\\
methylamine               &        -4.56 &        -4.93 (0.04) &        -6.44 (0.04)\\
methylether               &        -1.90 &        -1.30 (0.04) &        -1.43 (0.04)\\
methylethylsulfide        &        -1.50 &        -1.31 (0.06) &        -1.84 (0.06)\\
methylisopropylether      &        -2.01 &        -2.36 (0.06) &        -3.32 (0.06)\\
methylsulfide             &        -1.24 &        -1.06 (0.04) &        -1.20 (0.04)\\
n-butane                  &         2.08 &         2.26 (0.05) &         2.17 (0.05)\\
n-butanethiol             &        -0.99 &        -0.70 (0.06) &        -0.83 (0.06)\\
n-octane                  &         2.88 &         3.04 (0.09) &         2.51 (0.10)\\
octan-1-ol                &        -4.09 &        -4.72 (0.11) &        -5.12 (0.11)\\
p-cresol                  &        -6.14 &        -5.99 (0.07) &        -6.10 (0.08)\\
phenol                    &        -6.62 &        -5.21 (0.06) &        -5.99 (0.06)\\
propane                   &         1.96 &         2.36 (0.05) &         2.09 (0.05)\\
propanol                  &        -4.83 &        -4.83 (0.05) &        -5.57 (0.05)\\
propylamine               &        -4.40 &        -4.14 (0.06) &        -5.56 (0.05)\\
toluene                   &        -0.89 &        -0.27 (0.07) &        -0.63 (0.06)\\
trimethylamine            &        -3.24 &        -2.35 (0.05) &        -3.29 (0.05)\\
water                     &        -6.32 &        -6.80 (0.03) &        -7.56 (0.03)\\
\hline
  \end{tabular}
  \caption{Experimental, adQTB and classical values of the HFE (in kcal/mol)} for the dataset of organic molecules (alphabetical order, from A to L). Estimated standard errors are given in parenthesis.
  \label{fig:hfe_values}
\end{table}

\clearpage

\section{Radial distribution functions of liquid water using DeePMD}
\begin{figure}[h!]
    \centering
    \includegraphics[width=0.9\textwidth]{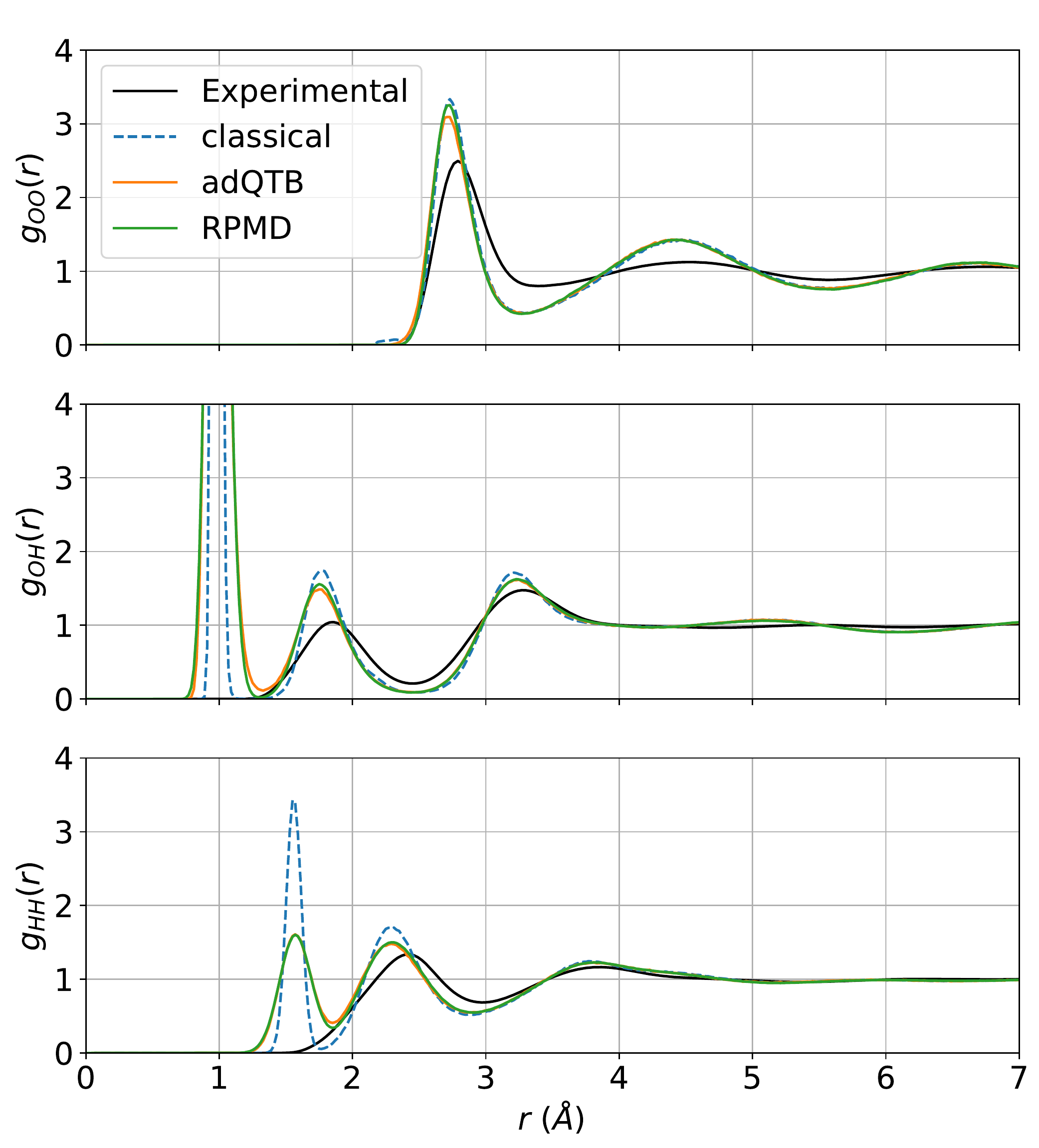}
    \caption{Radial distribution functions of liquid water at 300K computed using the DeePMD ML force field and simulated with classical dynamics (dashed), adQTB (solid orange) and RPMD (solid green). Experimental results from ref.~\cite{soper2013radial}.}
    \label{fig:gr_deepmd}
\end{figure}

Figure~\ref{fig:gr_deepmd} shows the radial distributions functions of liquid water at 300K simulated using the DeePMD neural network potential.

The DeePMD model was trained on path integral \textit{ab-initio} molecular dynamics (PI-AIMD) trajectories, at the PBE0-TS level (refs~\cite{hsinYu2019isotope,zhang2021phase}): (1) 100000 snapshots of PI-AIMD liquid water (192 atoms) at 1 bar and 300K (2) 20000 snapshots of PI-AIMD ice phase Ih (288 atoms) at 1 bar and 273K (3) 10000 snapshots of classical AIMD ice phase Ih at 1 bar and 330K and (4) 10000 snapshots of classical AIMD ice phase Ih at 2130 bar and 238K. We used 10\% of the data as validation set. The DeePMD model was trained using the DeePMD-kit package~\cite{zhang2018deep}. The DeePMD model architecture is composed of a (25, 50, 100) embedding net with a 18 neuron-size embedding sub-matrix, and a (240,240, 120, 60, 30, 10) fitting net. The cut-off radius was set to 6~\AA~with a smoothing cutoff of 0.5~\AA~and a two-body embedding descriptor. The final model is trained with $1.2\times10^{7}$ Adam steps.

We performed 500 ps of NVT simulation (at experimental density) for a cubic box of 1000 water molecules for both classical and adQTB MD, and a smaller box of 216 molecules for PIMD (with $\MM=32$ beads).

\bibliography{biblio}